\documentclass{article}
\usepackage[utf8]{inputenc}
\usepackage{authblk}

\usepackage[backend=biber,maxnames=3]{biblatex}
\usepackage{amsfonts}
\usepackage{amsmath}
\usepackage{amssymb}
\usepackage{amsthm}
\usepackage{enumitem}
\usepackage{graphicx}
\usepackage{enumitem}
\usepackage{geometry}
\usepackage{subfigure}
\usepackage[version=4]{mhchem}
 
\theoremstyle{plain}
\newtheorem{thm}{Theorem}[section]
\newtheorem{remark}[thm]{Remark}

\usepackage{xcolor}

\newcommand{\R}{\mathbb{R}}

\newcommand{\q}{\quad}

\newcommand{\p}{{\bf p}}
\renewcommand{\q}{{\bf q}}
\newcommand{\e}{{\bf e}}
\newcommand{\x}{{\bf x}}

\newcommand{\X}{{\bf X}}

\newcommand{\W}{{\bf W}}

\newcommand{\ud}{\mathrm{d}}

\newcommand{\fR}{f_+}
\newcommand{\fL}{f_-}
\newcommand{\J}{{\bf J}}
\newcommand{\n}{{\bf n}}

\newcommand{\Dtheta}{\Delta_\theta}

\title{Active Crowds}
\date{}

\author[1]{Maria Bruna}
\author[2]{Martin Burger}
\author[3]{Jan-Frederik Pietschmann}
\author[4]{Marie-Therese Wolfram}

\affil[1]{Department of Applied Mathematics and Theoretical Physics, University of Cambridge, Cambridge, CB3 0WA, UK}
\affil[2]{Department Mathematik, Friedrich-Alexander Universit\"at Erlangen-N\"urnberg
Cauerstrasse 11, 91058 Erlangen, Germany}
\affil[3]{Fakult\"at f\"ur Mathematik, Technische Universit\"at Chemnitz Reichenhainer Straße 41, Chemnitz, Germany}
\affil[4]{University of Warwick, Mathematics Institute, CV4 7AL Coventry, UK and Radon Institute for Computational And Applied Mathematics, Altenbergerstr. 69, 4040 Linz, Austria}

\begin{document}

\maketitle

\begin{abstract}
This chapter focuses on the mathematical modelling of active particles (or agents) in crowded environments. We discuss several microscopic models found in literature and the derivation of the respective macroscopic partial differential equations for the particle density. The macroscopic models share common features, such as cross diffusion or degenerate mobilities. We then take the diversity of macroscopic models to a uniform structure and work out potential similarities and differences. Moreover, we discuss boundary effects and possible applications in life and social sciences. This is complemented by numerical simulations that highlight the effects of different boundary conditions.
\end{abstract}

\section{Introduction}

The mathematical modelling of active matter has received growing interest recently, motivated by novel structures in physics and biology on the one hand (cf. \cite{bechinger2016active,desai2017modeling,elgeti2015physics,Gompper:2020dj,schmidt2019light,tailleur2008,zottl2016emergent}), but also active matter in a wider sense of agent systems like humans or robots (cf. \cite{bellomo2010modelling,Burger2016:sidestepping,Cirillo2020:ActivePassive,CPT2014,nelson2010microrobots,nava2018markovian,elamvazhuthi2019mean}). In many of these systems a key issue is the interplay of the particles' own activity with crowding effects, which leads to the formation of complex and interesting patterns. In this chapter we aim at unifying a variety of models proposed for such systems, discuss the derivation of macroscopic equations from different microscopic paradigms, and highlight some of their main properties.

First we would like to emphasise that the definition of active particles varies in the literature e.g., between condensed matter, life sciences and engineering and we that will adopt a generous point of view in this chapter.
We will discuss models for actual active matter systems, but also systems that might be considered passive (or force-driven) in the physics literature but have been used to model active matter systems. A common feature of these models is that the particles have a preferred direction of motion and can use energy to move there; the  preferred direction can however change in time. This setting includes models for multiple species (with fixed directions or biases) and systems with boundary conditions that impose steady currents. From a mathematical point of view we can distinguish whether models exhibit a gradient-flow structure (cf. \cite{AGS2008}) or not, respectively whether there are stationary solutions with vanishing flux. In the case of a gradient-flow structure there is a natural entropy-energy functional to be dissipated (cf. \cite{CJMTU2001}), in the other cases such functionals may increase (linearly) in time or it is not apparent what the correct choice of the functional is. We shall see in particular that gradient-flow structures are destroyed if particles change directions completely randomly, while there is an active transport in that direction. 

In the following we will consider models with a finite number of preferred directions or a continuum of it. While the former has been proposed and investigated in many applications like pedestrian dynamics or cell motility, continuum directions received far less attention in the mathematical literature. In the continuum case one can consider angular diffusion and derive an equation for the density of particles in the phase-space of spatial and angular variables. We discuss different microscopic models, either based on Brownian motions with hard sphere potentials or on lattice based models with size exclusion, which allow to derive appropriate macroscopic partial differential equations (PDEs) for the phase-space density. These PDEs have a rather similar structure - all have a nonlinear transport term and additional diffusion terms in space and angle (or possibly nonlocal diffusion for the latter). This general structure allows us to define a general entropy functional and investigate the long time behaviour.

This chapter is organised as follows: We present several microscopic models for active particles and their corresponding mean-field limits using different coarse-graining procedures in Section \ref{sec:active}. Section \ref{s:nonact} discusses the respective modelling approaches and limiting equations for externally activated particles (systems that would be considered passive in the physics literature). We then present a general formulation of all these models and state their underlying properties, such as energy dissipation or a possible underlying gradient-flow structure in Section \ref{sec:gen_structure}. The important role of boundary conditions on the behaviour of these systems is discussed in Section \ref{s:bc}. Section \ref{sec:applications} presents several examples of active and externally activated particle models in the life and social sciences. Numerical experiments illustrating the dynamics and behaviour of the respective models are presented in Section \ref{sec:numerics}.

Throughout this chapter we use the notion of particles or agents interchangeably. Furthermore we discuss the respective models on the line or in $\R^2$, with the obvious generalisation to three dimensions. We will keep the presentation informal, assuming that all functions satisfy the necessary requirements to perform all limits and calculations.

\section{Models for active particles} \label{sec:active}

Here we discuss microscopic models for active particles and their corresponding macroscopic kinetic models using different coarse-graining procedures. The key ingredients of active particles is that, in addition to their positions, they have an orientation that determines the self-propulsion direction. We subdivide these models into continuous, discrete or hybrid random walks depending on how the position and the orientation of each particle is represented. 

\subsection{Continuous random walks} \label{sec:continuous_walks}

We consider $N$ identical Brownian particles with free translational diffusion coefficient $D_T$ moving in a periodic box $\Omega \subset \mathbb R^2$ with unit area. Each particle has a position $\X_i(t)$ and an orientation $\Theta_i(t)$ with $t>0$, $i = 1, \dots, N$, that determines the direction $\e(\Theta_i) = (\cos \Theta_i, \sin \Theta_i)$ of self-propulsion at constant speed $v_0$. The orientation $\Theta_i$ also undergoes free rotational diffusion with diffusion coefficient $D_R$. 
In its more general form, particles interact through a pair potential $u(r, \varphi)$ which implies the total potential energy 
\begin{align}\label{e:U}
    U = \chi\sum_{1\le i<j \le N} u(|\X_i - \X_j|/\ell, |\Theta_i - \Theta_j|),
\end{align}
where $\chi$ and $\ell$ represent the strength and the range in space of the potential $u$, respectively. The coupled equations of motion are:
\begin{subequations}
	\label{sde_model}
\begin{align}
\label{sde_x}
	\ud \X_i &= \sqrt{2 D_T} \ud {\bf W}_i - \nabla_{\x_i} U\ud t + v_0 \e(\Theta_i) \ud t,\\
	\label{sde_angle}
	\ud \Theta_i &= \sqrt{2 D_R} \ud W_i - \partial_{\theta_i} U\ud t.
\end{align}
\end{subequations}
We note that isotropic pair potentials ($u = u(r)$ only) are more commonly used in the literature \cite{Bialke:2013gw,Keta:2021fl}. Equations \eqref{sde_model} are complemented with identically and independently distributed initial conditions, $(\X_i(0), \Theta_i(0)) \sim f_0(\x,\theta)$ and periodic boundary conditions on $\Upsilon = \Omega \times [0, 2\pi)$ (we will discuss alternative boundary conditions later in Section \ref{s:bc}).

The starting point for all is to define the joint probability density for $N$ particles evolving according to the SDEs \eqref{sde_model}, that is $F_N(\vec \xi, t)$ with $\vec \xi = (\xi_1, \dots, \xi_N)$ and $\xi_i = (\x_i, \theta_i)$. Using the Chapman--Kolmogorov equation, we obtain
\begin{equation} \label{N_eq}
	\partial_t F_N(\vec \xi, t) = \sum_{i=1}^N \nabla_{\x_i} \cdot \left[ D_T \nabla_{\x_i} F_N -v_0\e(\theta_i) F_N + \nabla_{\x_i} U(\vec \xi) F_N \right] + \partial_{\theta_i} \left[ D_R \partial_{\theta_i } F_N  + \partial_{\theta_i} U(\vec \xi) F_N\right],
\end{equation}
for $t \ge 0, \vec \xi \in \bar \Upsilon$, where $\bar \Upsilon \subseteq \Upsilon^N$ is the domain of allowed configurations (more on this below). 

The goal is to obtain a macroscopic model for the one-particle density $f(\xi, t)$, that we can describe by picking the first particle since all particles are identical, i.e.,
\begin{align}
	f(\xi_1,t) = \int_{\Upsilon^N} F_N(\vec \xi) \delta(\xi - \xi_1) \ud \vec \xi.
\end{align}
To this end, keeping in mind all the particles are indistinguishable, we integrate \eqref{N_eq} with respect to $\xi_2, \dots, \xi_N$. Using periodicity, all the terms for $i\ge 2$ in the right-hand side of \eqref{N_eq} vanish, and we are left with:
\begin{subequations}
	\label{general_macro}
\begin{equation} \label{1_eq}
	\partial_t f(\xi_1, t) = \nabla_{\x_1} \cdot \left[ D_T \nabla_{\x_1} f - v_0\e(\theta_1) f + {\bf U}_T(\xi_1,t) \right] + \partial_{\theta_1} \left[ D_R \partial_{\theta_1 } f  + U_R(\xi_1,t)\right],
\end{equation}
with 
\begin{align} \label{interaction_G}
&\begin{aligned}
	{\bf U}_T(\xi_1,t) &= \chi \int_{\Upsilon^{N-1}} F_N(\vec \xi, t) \sum_{i=2}^N \nabla_{\x_1} u(|\x_1-\x_i|/\ell,|\theta_1 -\theta_i|) \ud \xi_2, \dots, \ud \xi_N \\
	&= \chi (N-1) \int_{\Upsilon} F_2(\xi_1,\xi_2,t) \nabla_{\x_1} u(|\x_1-\x_2|/\ell,|\theta_1 -\theta_2|) \ud \xi_2,
	\end{aligned}\\
&\begin{aligned}
	U_R(\xi_1,t) &= \chi \int_{\Upsilon^{N-1}} F_N(\vec \xi, t) \sum_{i=2}^N \partial_{\theta_1} u(|\x_1-\x_i|/\ell,|\theta_1 -\theta_i|) \ud \xi_2, \dots, \ud \xi_N \\
	&= \chi (N-1) \int_{\Upsilon} F_2(\xi_1,\xi_2,t) \partial_{\theta_1} u(|\x_1-\x_2|/\ell,|\theta_1 -\theta_2|) \ud \xi_2,
	\end{aligned}
\end{align}
using that particles are undistinguishable, where $F_2$ is the two-particle density
$$
F_2(\xi_1,\xi_2,t) = \int_{\Upsilon^{N-2}} F_N(\vec \xi,t) \ud \xi_3 \dots \ud \xi_N.
$$
\end{subequations}

Depending on the scalings $\chi, \ell$ of the interaction potential $u$, we can expect different macroscopic limits of \eqref{sde_model}. On the one end we can consider long-range weak repulsive interactions, and obtain a mean-field limit equation. On the other extreme, we can consider short and strong repulsive interactions (even hard-core interactions such as $u(r) = +\infty$ if $r<1$, and 0 otherwise), which lead to local nonlinear PDE models.

\subsubsection{Mean-field scaling}
The mean-field scaling corresponds to $\chi = 1/N$ and $\ell = O(1)$ so that we have a weak and long range interaction in the limit of $N\to \infty$. In this limit, one expects propagation of chaos leading to 
$$
F_2(\xi_1, \xi_2, t) \approx f(\xi_1,t) f(\xi_2, t).
$$
Substituting this into \eqref{general_macro} we arrive at
\begin{subequations}
	\label{mfa_model}
\begin{equation}
	\label{1_mfa}
	\partial_t f(\xi_1, t) = \nabla_{\x_1} \cdot \left[ D_T \nabla_{\x_1} f -v_0 \e(\theta_1) f +  f \nabla_{\x_1} \mathcal U \right] + \partial_{\theta_1} \left[ D_R \partial_{\theta_1 } f  + \partial_{\theta_1} \mathcal U\right],
\end{equation}
with interaction term, taking $N\to \infty$,
\begin{equation} \label{U_functional}
	\mathcal U(f) = \int_\Upsilon f(\xi_2,t) u(|\x_1-\x_2|/\ell,|\theta_1 -\theta_2|) \ud \xi_2.
\end{equation}
\end{subequations}

In the case of an isotropic interaction potential, the term $\partial_{\theta_1} \mathcal U$ in \eqref{1_mfa} drops and the interaction term \eqref{U_functional} can be simplified to
\begin{equation} \label{U_functional_red}
	\mathcal U(f) = \int_\Omega \rho(\x_2,t) u(|\x_1-\x_2|/\ell) \ud \x_2,
\end{equation}
where $\rho$ is the spatial (macroscopic) density
\begin{align}\label{local_rho}
	\rho(\x,t) &= \int_0^{2\pi} f(\x,\theta,t) ~\ud\theta.
\end{align}
The spatial density describes the probability that a particle is at position $\x$ at time $t$ irrespective of its orientation. We obtain the following equation for $\rho$ by integrating \eqref{1_mfa} with the potential \eqref{U_functional_red} with respect to $\theta \in [0, 2 \pi)$ and using periodicity:
\begin{equation}
	\label{mfa_rho}
	\partial_t \rho(\x_1, t) = \nabla_{\x_1} \cdot \left[ D_T \nabla_{\x_1} \rho -v_0 \p +  \rho \nabla_{\x_1} \mathcal U \right].
\end{equation}
This equation is not closed as it depends on the polarisation $\p$ (also known as the order parameter):
\begin{align}\label{polarisation}
	\p(\x,t) = \int_0^{2\pi} \e(\theta) f(\x,\theta,t) ~\ud\theta.
\end{align}
The polarisation gives the average orientation of particles at position $\x$ at any given time $t$. 

\subsubsection{Excluded-volume interactions}\label{sec:evi}
Excluded-volume interactions are very common in biological applications and arise from the impenetrability between cells, bacteria, animals, etc. These are very strong and short-range interactions, whereby an individual only interacts locally in the range of its body size. For these reasons, the mean-field scaling is not suitable to model such interactions, which are often modelled using singular short-range potentials ($\ell \ll 1$ in \eqref{e:U}) or even hard-core potentials. Examples of interactions potentials used in the literature to model excluded-volume interactions include inverse power-law potentials (such as the Lennard-Jones potentials), exponential potentials (e.g., the Morse potential), or the Yukawa potential.

The following model, proposed by \cite{Speck:2015um}, includes excluded-volume interactions via a short-range interaction potential $u(r)$:
\begin{align}\label{model2}
\partial_t f + \nabla \cdot ( v_e(\rho) f \e(\theta)) &= D_e(\phi) \Delta f + D_R \partial_{\theta}^2 f,
\end{align}
where $f = f(r, \theta, t)$,  $D_e(\phi)$ is an effective diffusion depending on how crowded the system is (given by $\phi$),  and  $v_e = v_0 (1 - \phi \rho)$ is a nonlinear effective speed. The hydrodynamic equations for the spatial density $\rho$ and the polarisation $\p$ are obtained by integrating \eqref{model2},
\begin{align} \label{model2:rho}
	\partial_t \rho + \nabla \cdot ( v_e(\rho) \p) &= D_e(\phi) \Delta \rho,\\
	\partial_t \p + \nabla P(\rho) & = D_e(\phi) \Delta \p - \p,
\end{align}
with so-called pressure $P(\rho) = v_e(\rho) \rho /2$. 
This model displays a motility-induced phase transition \cite{Speck:2015um}: at low densities ($\phi \rho$ small), the effective swimming speed is close to the free speed $v_0$, whereas at high densities, the effective swimming goes to zero. The result is a phase separation, which regions of high density where particles are trapped and do not move, and very dilute areas with fast speeds. This is shown via a linear stability analysis as well as numerical simulations of the microscopic system using the repulsive Weeks-Chandler-Andersen (WCA) potential (which corresponds to a truncated and shifted upwards Lennard--Jones potential). Through an adiabatic approximation, they cast equation \eqref{model2:rho} into a gradient-flow of an effective free energy of the form of a conventional Ginzburg-Landau function. According to \cite{Speck:2015um}, this is consistent with   ``the mapping of active phase separation onto that of passive fluids with attractive interactions through a global effective free energy''.

An alternative derivation of a macroscopic model for active Brownian particles is considered in \cite{Bruna:2021tb} using the hard-core interaction potential, $u(r) = +\infty$ for $r<\epsilon$ and 0 otherwise. In this case, the microscopic model changes from \eqref{sde_model} to
\begin{subequations}
	\label{sde_model_hc}
\begin{align}
\label{sde_x_hc}
	\ud \X_i &= \sqrt{2 D_T} \ud {\bf W}_i + v_0 \e(\Theta_i) \ud t, \qquad |\X_i -\X_i| > \epsilon, \forall j\ne i, \\
	\label{sde_angle_hc}
	\ud \Theta_i &= \sqrt{2 D_R} \ud W_i.
\end{align}
\end{subequations}
This represents particles as hard disks of diameter $\epsilon$: particles only sense each other when they come into contact, and they are not allowed to get closer than $\epsilon$ to each other (mutual impenetrability condition). In comparison with the mean-field scaling, here instead the scaling is $\chi = 1, \ell = \epsilon \ll 1$ so that each particle only interacts with the few particles that are within a distance $O(\epsilon)$, the interaction is very strong. Using the method of matched asymptotics, from \eqref{sde_model_hc} one obtains to order $\phi$ the following model:
\begin{align}\label{model3}
\partial_t f + v_0\nabla \cdot \left[ f (1-\phi \rho) \e(\theta) + \phi \p f\right] &= D_T \nabla \cdot \left[ (1- \phi \rho) \nabla f + 3 \phi f \nabla  \rho \right]  + D_R \partial_{\theta}^2 f.
\end{align}
Here $\phi$ is the effective occupied area $\phi = (N-1)\epsilon^2 \pi/2$. Model \eqref{model3} is obtained formally in the limit of $\epsilon$ and $\phi$ small. 
Note that this equation is consistent with the case $N=1$: if there is only one particle, then $\phi = 0$ and we recover a linear PDE (no interactions). The equation for the spatial density is
\begin{equation}
\partial_t \rho + v_0 \nabla \cdot  \p  = D_T \nabla \cdot \left[ (1+ 2\phi \rho) \nabla \rho \right],
\end{equation}
which indicates the collective diffusion effect: the higher the occupied fraction $\phi$, the higher the effective diffusion coefficient.  
We note that, due to the nature of the excluded-volume interactions, models \eqref{model2} and \eqref{model3} are obtained via approximations (closure at the pair correlation function and matched asymptotic expansions, respectively) and no rigorous results are available. A nice exposition of the difficulties of going from micro to macro in the presence of hard-core non-overlapping constraints is given in \cite{Maury:2011eu}. In particular, they consider hard-core interacting particles in the context of congestion handling in crowd motion. In contrast to \eqref{sde_model_hc}, the dynamics involve only position and are deterministic. Collisions can then be handled via the projection of velocities onto the set of feasible velocities. In \cite{Maury:2011eu} they do not attempt to derive a macroscopic model from the microscopic dynamics but instead propose a PDE model for the population density $\rho(\x,t)$ that expresses the congestion assumption by setting the velocity to zero whenever $\rho$ attains a saturation value (which they set to one). 

\subsection{Discrete random walks}\label{sec:discrete}

\begin{figure}[h!]
    \centering
    \includegraphics[width=0.8\textwidth]{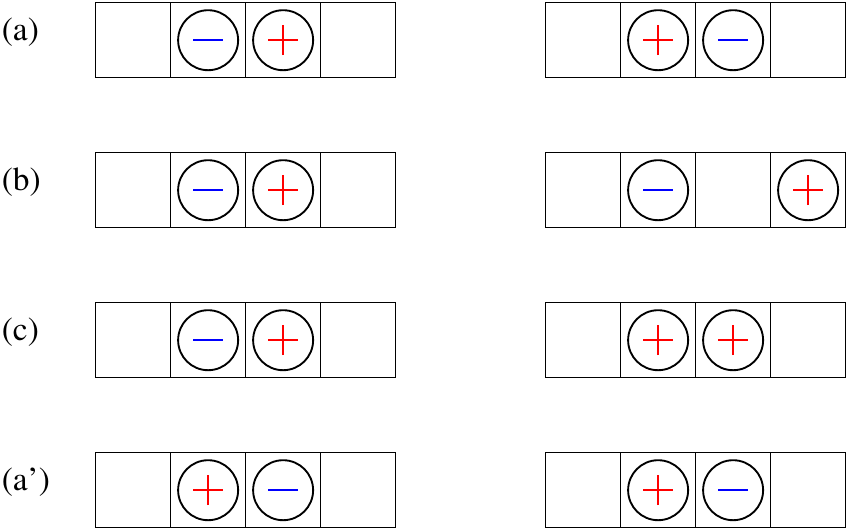}
    \caption{Sketch illustrating the update steps for $+$ (right moving) and $-$ (left moving) particles outlined in Section \ref{sec:discrete}. The left column shows the initial setup, the right one the configuration after a single time step.}
    \label{fig:discrete1d}
\end{figure}

Next we discuss fully discrete models for active particles with size exclusion effects. We start by considering a 
simple exclusion model for active particles on a one-dimensional lattice, which has been investigated in \cite{kourbane2018exact}. The brief description of the microscopic lattice model is as follows: $N$ particles of size $\epsilon$ evolve on a discrete ring of $1/\epsilon$ sites, with occupancy $\phi = \epsilon N \le 1$. Each lattice is occupied by at most one particle (thus modelling a size exclusion), and particles can either be moving left ($-$ particles) or right ($+$ particles). The configuration can be represented using occupation numbers $\sigma_i$ at site $i$ with values in $\{-1, 0, 1\}$. 
The dynamics combine three mechanisms: 
\begin{enumerate}[label=(\alph*)]
	\item Diffusive motion: for each bond $(i, i+1)$, $\sigma_i$ and $\sigma_{i+1}$ are exchanged at rate $D_T\backslash \epsilon^2$.
	\item Self-propulsion and size exclusion: for each bond $(i, i+1)$, a $+$ particle in $i$ jumps to $i+1$ if $\sigma_{i+1} = 0$; or a $-$ particle in $i+1$ jumps to $i$ if $\sigma_i = 0$, both at rate $\epsilon v_0$.
	\item Tumbling: particles switch direction $\sigma_i \to - \sigma_i$ at rate $\epsilon^2 \lambda$,
\end{enumerate}
see Figure \ref{fig:discrete1d} for an illustration of these effects.
Rescaling space and time as $\epsilon i$ and $\epsilon^2 t$ respectively, and a smooth initial condition, the macroscopic equations can be derived exactly as \cite{kourbane2018exact}
\begin{align} \label{model_lattice1D}
\begin{aligned}
\partial_t \fR + v_0\partial_x [\fR (1-\phi \rho)] &= D_T\partial_{xx} \fR + \lambda (\fL - \fR), \\
\partial_t \fL - v_0\partial_x [\fL (1-\phi \rho)] &= D_T\partial_{xx} \fL + \lambda (\fR - \fL),
\end{aligned}
\end{align}
where $\fR$ and $\fL$ are the probability densities corresponding to the $+$ and $-$ particles, respectively, and $\rho = \fR + \fL$. 
Introducing the number densities
\begin{align}\label{number_densities}
r(\x,t) = N f_+(\x,t), \qquad b(\x,t) = N f_-(\x,t),
\end{align}
which integrate to $N_1$ and $N_2$ respectively, we can rewrite \eqref{model_lattice1D} as 
\begin{align}\label{model_lattice1D_number_densities}
\begin{aligned}
\partial_t r + v_0\partial_x [r (1-\bar \rho)] &= D_T\partial_{xx} r + \lambda (b - r), \\
\partial_t b - v_0\partial_x [b (1- \bar\rho)] &= D_T\partial_{xx} b + \lambda (r - b),
\end{aligned}
\end{align}
with $\bar \rho = \epsilon (r + b)$.
One can also consider the same process in higher dimensions with a finite set of orientations $\e_k, k = 1, \dots, m$. The most straightforward generalisation of \eqref{model_lattice1D} is to consider a two-dimensional square lattice with $m=4$ directions, namely $\e_1 = (1,0), \e_2 = (0,1), \e_3 = (-1,0), \e_4 = (0,-1)$ (see Fig.~1 in \cite{kourbane2018exact}). In this case, the configuration would take five possible values, $\sigma_i= \{-1, -i, 0, i, 1\}$ and the resulting macroscopic model would consist of a system of four equations for the densities of each subpopulations
\begin{equation} \label{lattice2D}
	\partial_t f_k + v_0\nabla \cdot [f_k (1-\phi \rho) \e_k] = D_T \Delta f_k + \lambda (f_{k+1} + f_{k-1} - 2 f_k), \qquad k = 1, \dots,4
\end{equation}
where now $\phi = \epsilon^2 N$, $f_k(\x,t)$ stands for the probability density of particles going in the $\e_k$ direction, and $\rho = \sum_k f_k$. Periodicity in angle implies that $f_5=f_1, f_{-1} = f_4$. 

Note how the model in \cite{kourbane2018exact} differs from an asymmetric simple exclusion processes (ASEP) in that particles are allowed to swap places in the diffusive step (see (a) above). As a result, the macroscopic models \eqref{model_lattice1D} and \eqref{lattice2D} lack any cross-diffusion terms. We can also consider an actual ASEP process, in which simple exclusion is also added to the diffusive step, that is, point (a) above is replaced by
\begin{enumerate}[label=(\alph*')]
	\item Diffusive motion: a particle in $i$ jumps to $i+ 1$ at rate $D_T\backslash \epsilon^2$ if $\sigma_{i+ 1} = 0$ (and similarly to $i-1$).
\end{enumerate}
In this case, the resulting macroscopic model is
\begin{equation} \label{ASEP_2D}
	\partial_t f_k + v_0\nabla \cdot [ f_k (1- \phi \rho) \e_k] = D_T \nabla \cdot [(1- \phi \rho) \nabla f_k + \phi f_k \nabla \rho] + \lambda (f_{k+1} + f_{k-1} - 2 f_k), \qquad k = 1, \dots,4.
\end{equation}

\subsection{Hybrid random walks}\label{sec:hybrid}

In the previous two subsections we have discussed models that consider both the position and the orientation as continuous, or discrete. Here we discuss hybrid random walks, that is, when positions are continuous and orientations finite, or vice-versa.

The first hybrid model we consider is an active exclusion process whereby the orientation is a continuous process in $[0, 2 \pi)$ evolving according to a Brownian motion with diffusion $D_R$, \eqref{sde_angle}, while keeping the position evolving according to a discrete asymmetric exclusion process (ASEP) \cite{Bruna:2021tb}. The advantage of this approach is to avoid the anisotropy imposed by the underlying lattice. 
Here we present the model in two-dimensions so that we can compare it to the models presented above.  

\begin{figure}[h!]
    \centering
    \includegraphics{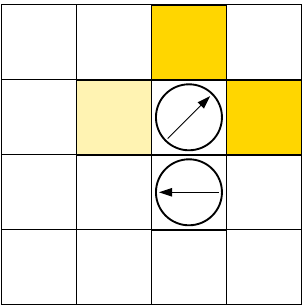}
    \caption{Sketch of the 2D hybrid random walk  outlined in Section \ref{sec:hybrid}. The arrow within each particle corresponds to its orientation, the colour of the neigneighbouringhboring sites relate to $\pi_k$. The darker the color, the greater the likelihood to jump into the cell.}
    \label{fig:hybrid2d}
\end{figure}

We consider a square lattice with spacing $\epsilon$ and orientations $\e_k, k=1, \dots, 4$ as given above. A particle at lattice site $\x$ can jump to neighbouring sites $\x+\epsilon \e_k$ if the latter is empty at a rate $\pi_k(\theta)$ that depends on its orientation $\theta$, namely 
\begin{equation*}
	\pi_k(\theta) = \alpha_\epsilon \exp(\beta_\epsilon \e(\theta) \cdot \e_k),
\end{equation*}
where $\alpha_\epsilon = D_T /\epsilon^2$ and $\beta_\epsilon = v_0 \epsilon/(2 D_T)$. Therefore, the diffusive and self-propulsion mechanisms in \eqref{model_lattice1D} are now accounted for together: jumping in the direction opposite to your orientation reduces the rate to $\sim \alpha_\epsilon(1-\beta_\epsilon)$, whereas the there is a positive bias $\sim \alpha_\epsilon(1+\beta_\epsilon)$ towards jumps in the direction pointed to by $\e(\theta)$, see Figure \ref{fig:hybrid2d} The tumbling (point 3 above) is replaced by a rotational Brownian motion. 

Taking the limit $\epsilon\to 0$ while keeping the occupied fraction $\phi =N \epsilon^2$ finite one obtains the following macroscopic model for $f = f(\x,\theta,t)$
\begin{align} \label{model_hy}
\partial_t f + v_0\nabla \cdot [ f (1- \phi \rho) \e(\theta)] &= D_T \nabla \cdot ( (1- \phi \rho) \nabla f + \phi f \nabla \rho)  + D_R \partial_{\theta}^2 f.
\end{align}
This model can be directly related to the fully discrete model \eqref{ASEP_2D}: they are exactly the same if one considers \eqref{ASEP_2D} as the discretised-in-angle version of \eqref{model_hy} by identifying 
$$
D_R \partial_\theta^2 f_k \approx D_R \frac{f_{k+1} + f_{k-1} - 2 f_k}{(2\pi/m)^2}, 
$$
that is, $\lambda = D_R m^2/(2 \pi^2)$, where $m$ is the number of orientations in the fully discrete model.

The other possible hybrid model is to consider a continuous random walk with interactions in space \eqref{sde_x}, while only allowing a finite number of orientations,  $\Theta_i \in \{\theta_1, \dots, \theta_m\}$. In its simplest setting, we can consider that $\theta_k$ are equally spaced in $[0, 2\pi)$ and a constant switching rate $\lambda$ between the neighbouring angles. The $N$ particles evolve according to the stochastic model:
\begin{subequations} \label{hybrid2}
\begin{align}
\label{hybrid_x}
\ud \X_i &= \sqrt{2 D_T} \ud {\bf W}_i - \nabla_{\x_i} U\ud t + v_0 \e(\Theta_i) \ud t,\\
\label{hybrid_angle}
\Theta_i &= \{\theta_k\}_{k=1}^m, \quad \theta_k \xrightarrow{\ \lambda \ } \theta_{k+ 1}\pmod{2\pi}, \quad \theta_k \xrightarrow{\ \lambda \ } \theta_{k- 1}\pmod{2\pi}.
\end{align}
\end{subequations}
If we assume excluded-volume interactions through a hard-core potential, the resulting model is \cite{Wilson:2018fg}
\begin{equation}\label{model_wilson}
	\partial_t f_k + v_0\nabla \cdot \left[ f_k (1-\phi \rho) \e_k + \phi \p f_k\right] = D_T \nabla \cdot \left[ (1- \phi \rho) \nabla f_k + 3 \phi f_k \nabla  \rho \right]  + \lambda \left ( f_{k+1} + f_{k-1} -2 f_k \right),
\end{equation}
where $\rho = \sum_{k=1}^m f_k$, $\p = \sum_{k=1}^m f_k \e(\theta_k)$, and $\e_k = \e(\theta_k)$. 
The density $f_k(\x,t)$ represents the probability of finding a particle at position $\x$ at time $t$ with orientation $\theta_k$ (naturally, we identify $f_{m+1} = f_1$ and $f_{-1} = f_m$). Here $\phi = (N-1) \epsilon^2 \pi/2$ represents the effective excluded region as in \eqref{model3}. We note how this model is consistent with the continuous model \eqref{model3}, in that if we had discretised angle in \eqref{model3} we would arrive at the cross-diffusion reaction model \eqref{model_wilson}. 

A variant of the hybrid model \eqref{hybrid2} is to allow for jumps to arbitrary orientations instead of rotations of $2\pi/m$, namely, from $\theta_k$ to $\theta_{j}\pmod{2\pi}$, $j\ne k$, at a constant rate $\lambda$ independent of the rotation. This is a convenient way to model the tumbles of a run-and-tumble process, such as the one used to describe the motion of \emph{E. Coli} \cite{Berg:1993ug}, see also Section \ref{sec:biological_transport}. In this case, the reaction term in \eqref{model_wilson} changes to 
\begin{equation}\label{hybrid3}
	\partial_t f_k + v_0\nabla \cdot \left[ f_k (1-\phi \rho) \e(\theta_k) + \phi \p f_k\right] = D_T \nabla \cdot \left[ (1- \phi \rho) \nabla f_k + 3 \phi f_k \nabla  \rho \right]  + \lambda  \sum_{j \ne k } \left( f_{j} - f_k \right).
\end{equation}

We may generalise the jumps in orientation by introducing a turning kernel $T(\theta, \theta')$ as the probability density function for a rotation from $\theta'$ to $\theta$. That is, if $\Theta_i(t)$ is the orientation of the $i$th particle at time $t$ and the jump occurs at $t^*$, 
$$
T(\theta, \theta') \ud \theta = \mathbb P\left(\left \{ \theta \le \Theta_i(t^*_+) \le \theta + \ud \theta \ |  \ \Theta_i(t^*_-) = \theta' \right \}\right).
$$
Clearly for mass conservation we require that $\int T(\theta, \theta') \ud \theta = 1$. 
The jumps may only depend on the relative orientation $\theta - \theta'$ in the case of a homogeneous and isotropic medium, in which case $T(\theta, \theta') \equiv  T(\theta-\theta')$. This is the case of the two particular examples above: in \eqref{model_wilson}, the kernel is 
$$
T(\theta, \theta') = \frac{1}{2} \left [ \delta(\theta -\theta'- \Delta) + \delta(\theta-\theta' + \Delta) \right], \qquad \Delta = \frac{2\pi}{m},
$$
whereas the rotation kernel in \eqref{hybrid3} is
$$
T(\theta, \theta') = \frac{1}{m-1} \sum_{k = 1}^{m-1} \delta(\theta -\theta' + k\Delta), \qquad  \Delta = \frac{2\pi}{m},
$$
where the argument of the delta function is taken to be $2\pi$-periodic. If the turning times $t^*$ are distributed according to a Poisson process with intensity $\lambda$, the resulting macroscopic model for the phase density $f = f(\x,\theta,t)$ with a general turning kernel $T$ becomes
\begin{equation}\label{run-tumble-cont}
	\partial_t f + v_0\nabla \cdot \left[ f (1-\phi \rho) \e(\theta) + \phi \p f\right] = D_T \nabla \cdot \left[ (1- \phi \rho) \nabla f + 3 \phi f \nabla  \rho \right]  -\lambda f +  \lambda  \int_0^{2\pi} T(\theta,\theta') f(\x,\theta',t) \ud \theta '.
\end{equation}
We note that the microscopic process associated with \eqref{run-tumble-cont} is continuous (and not hybrid) if the support of $T$ has  positive measure.

\section{Models for externally activated particles}\label{s:nonact}

In this section we go from active to passive particles and consider models with time reversal at the microscopic level. As mentioned in the introduction, the defining factor of active matter models is the self-propulsion term, which makes them out-of-equilibrium. Mathematically, this can be expressed by saying that even the microscopic model lacks a gradient-flow structure (either due to the term $\e(\theta)$ in the transport term, see \eqref{model3}, or the reaction terms in \eqref{model_lattice1D}, \eqref{lattice2D},  see Section \ref{sec:gen_structure}).
 
In the previous section we have seen the role the orientation $\theta$ plays. If it is kept continuous, the resulting macroscopic model is of kinetic type for the density $f(\x,\theta,t)$. If instead only a fixed number $m$ of orientations are allowed, then these define a set of $m$ species, whereby all the particles in the same species have the same drift term. This motivates the connection to cross-diffusion systems for passive particles, which are obtained by turning off the active change in directions in the models of Section \ref{sec:active} and look at the resulting special cases. This is a relevant limit in many applications, such as in pedestrian dynamics (see Section \ref{sec:pedestrian}).
Once the orientations are fixed, we are left with two possible passive systems: either originating from a spatial Brownian motion or a spatial ASEP discrete process. 

\subsection{Continuous models}
The starting point is the microscopic model \eqref{sde_model} taking the limit $D_R \to 0$. We could still keep the interaction potential as depending on the relative orientations, which would lead to different self- and cross-interactions (which might be useful in certain applications). Here for simplicity we assume interactions are all the same regardless of the orientations:
\begin{subequations}
	\label{sde_model_passive}
\begin{align}
\ud \X_i &= \sqrt{2 D_T} \ud {\bf W}_i - \nabla_{\x_i} U\ud t + v_0 \e(\Theta_i) \ud t,\\
\Theta_i(t) &= \theta_k, \qquad \text{if } i \in \mathcal I_k, \qquad k = 1,\dots, m,
\end{align}
\end{subequations}
where $\mathcal I_k$ is the set of particles belonging to species $k$. The number of particles in each species is $|\mathcal I_k| = N_k$.

The mean-field limit of \eqref{sde_model_passive} is given by (taking $N = \sum_k N_k \to \infty$ as in \eqref{1_mfa})
\begin{equation}
	\label{mfa_passive}
	\partial_t f_k(\x, t) = \nabla_{\x} \cdot \left[ D_T \nabla_\x f_k -v_0 \e(\theta_k) f_k +  f_k \nabla_{\x} (u \ast \rho)  \right],
\end{equation}
and $\rho(\x,t) = \sum_k f_k$. For consistency with the active models, here we do not take $f_k$ to be probability densities but to integrate to the relative species fraction, whereas as before the total density $\rho$ has unit mass:
\begin{equation} \label{normalisation_passive}
	\int_\Omega f_k(\x,t)  \ud \x = \frac{N_k}{N},\qquad \int_\Omega \rho(\x,t) \ud \x = 1.
\end{equation}
Thus $f_k = f_k(\x,t)$ describes the probability that a particle is at position $\x$ at time $t$, \emph{and} is in the $\mathcal I_k$ set.

The microscopic model \eqref{sde_model_passive} with the interaction term $U$ replaced by a hard-core potential for particles with diameter $\epsilon$ can be dealt with via the method of matched asymptotics. In this case, the resulting cross-diffusion model is
\begin{align}\label{eq:MF_cross_diff_sys}
	\partial_t f_k + v_0\nabla \cdot \left[ f_k \e_k  + \phi_{kl}(\e_l - \e_k) f_k f_l \right] &= D_T \nabla \cdot \left[ (1 + \phi_{kk} f_k) \nabla f_k + \phi_{kl} (3 f_k \nabla f_l - f_l \nabla f_k) \right], \qquad l \ne k,
\end{align}
where $\phi_{kk} = (N_k -1) N/N_k \epsilon^2 \pi$, $\phi_{kl} = N\epsilon ^2 \pi /2$ for $l\ne k$, and $f_k(\x,t)$ are defined as above. 
This model was first derived in \cite{Bruna:2012wu} for just two species but in a slightly more general context, whereby particles many have different sizes and diffusion coefficients
(also, note that in \cite{Bruna:2012wu}, \eqref{eq:MF_cross_diff_sys} appears written in terms of probability densities). 
Equation \eqref{eq:MF_cross_diff_sys} can be directly related to model \eqref{model_wilson} with $\lambda = 0$ if in both models we assume $N_k$ large enough such that $N_k - 1 \approx N_k, N-1 \approx N$: 
\begin{equation}\label{cross-diff_num_den}
	\partial_t f_k + v_0\nabla \cdot \left[ f_k (1-\phi \rho) \e(\theta_k) + \phi \p f_k\right] = D_T \nabla \cdot \left[ (1- \phi \rho) \nabla f_k + 3 \phi f_k \nabla  \rho	 \right],
\end{equation}
where $\phi = N \epsilon^2 \pi/2$, $\rho = \sum_k f_k$, and $\p =\sum_k f_k \e(\theta_k)$. Model \eqref{cross-diff_num_den} is the cross-diffusion system for red and blue particles studied in \cite{BBRW2017} in disguise. First, set the number of species to $m=2$ and define the number densities
\begin{align}\label{number_densitiesb}
r(\x,t) = N f_1(\x,t), \qquad b(\x,t) = N f_2(\x,t),
\end{align}
which integrate to $N_1$ and $N_2$ respectively. Then define the potentials $V_r = - (v_0/D_T) \e(\theta_1) \cdot \x$ and $V_b = - (v_0/D_T) \e(\theta_2) \cdot \x$. In terms of these new quantities, system \eqref{cross-diff_num_den} becomes
\begin{subequations}\label{e:aa_cross_sys}
\begin{align}
\partial_t r &= D_T \nabla \cdot \left[ (1+ 2\varphi r - \varphi b) \nabla r  + 3\varphi r \nabla b  + r \nabla V_r + \varphi r b \nabla (V_b - V_r)  \right],\\
\partial_t b &= D_T \nabla \cdot \left[ (1+2\varphi b - \varphi r) \nabla b  + 3\varphi  b \nabla r  + b \nabla V_b + \varphi  r b \nabla (V_r -  V_b)  \right],
\end{align}
\end{subequations}
where $\varphi = \epsilon^2 \pi/2$.
This is exactly the cross-diffusion system for particles of the same size and diffusivity studied  in \cite{BBRW2017} for $d=2$ (see Eqs.  (11) in \cite{BBRW2017}).\footnote{We note a typo in \cite{BBRW2017}: the coefficient $\beta$ below system (11) should have read $\beta = (2d-1)\gamma$.} 

\subsection{Discrete models}\label{sec:discrete_passive}

In this category there are discrete processes in space without changes in orientations. The most well-known model in the context of excluded-volume interactions is ASEP, which was used above in combination of either continuous change in angle, see \eqref{model_hy}, or discrete jumps, see \eqref{ASEP_2D}. We obtain the corresponding passive process by either setting $D_R$ or $\lambda$ to zero, respectively. The resulting model in either case is
\begin{align} \label{model3_passive}
\partial_t f_k + v_0\nabla \cdot [ f_k (1- \phi \rho) \e(\theta_k)] &= D_T \nabla \cdot [(1- \phi \rho) \nabla f_k + \phi f_k \nabla \rho], \qquad k = 1, \dots, m,
\end{align}
where $f_k$ satisfy \eqref{normalisation_passive} as before, and $\phi = N \epsilon^2$. We notice three differences with its continuous passive counterpart \eqref{cross-diff_num_den}: in the latter, the effective occupied fraction $\phi$ has a factor of $\pi/2$, the coefficient in the cross-diffusion term $f_k \rho$ has a factor of   three, and the transport term has an additional nonlinearity that depends on the polarisation.
The cross-diffusion system \eqref{model3_passive} was derived in \cite{Simpson:2009gi} and analysed in \cite{Burger:2010gb} for two species ($m=2$). Specifically, if we introduce the number densities $r, b$ and general potentials $V_r, V_b$ as above, it reads
\begin{subequations}
\label{eq:MF_cross_diff}
\begin{align}
\partial_t r &= D_T \nabla \cdot \left[(1-\bar \rho) \nabla r +  r \nabla \bar \rho + r (1- \bar \rho) \nabla V_r \right]\\
\partial_t b &= D_T \nabla \cdot \left[(1-\bar \rho) \nabla b +  b \nabla \bar \rho + b (1-\bar \rho) \nabla V_b \right],
\end{align}
\end{subequations}
where $\bar \rho = \epsilon^2 (r+b) = \epsilon^2 (N_1 f_1 + N_2 f_2)$ (compare with (3.7)-(3.8) in \cite{Burger:2010gb}).\footnote{In the system (3.7)-(3.8) of \cite{Burger:2010gb}, $r$ and $b$ are volume concentrations, thus having a factor of $\epsilon^2$ compared to those used in \eqref{eq:MF_cross_diff}, and the diffusivities of the two species are $1$ and $D$ instead of $D_T$ for both.} 

\section{General model structure}\label{sec:gen_structure}

We now put the models presented in the previous sections into a more general picture. We assume that $f = f(\x, \theta, t)$, where $\theta$ is a continuous variable taking values in $[0, 2\pi)$ or a discrete variable taking values $\theta_k$ for $k = 1, \dots, m$ (ordered increasingly on $[0,2\pi)$). For now on, we consider the density rescaled by $\phi$ instead of a probability density. This implies that $\phi$ disappears from the equations and enters the mass condition as $\iint f = \int \rho = \phi$. In the latter case we shall also use the notation $f_k(\x,t) = f(\x,\theta_k,t).$ We also recall the definition of the space density $\rho$ and the polarisation  $\mathbf{p}$:
 \begin{align*}
     \rho(\x,t) = \int_0^{2\pi} f(\x, \theta, t ) \, \ud \mu(\theta) \quad \text{and} \quad \mathbf{p}(\x,t) = \int_0^{2\pi} \e(\theta) f(\x, \theta, t) \, \ud \mu(\theta),
 \end{align*}
 where the integral in $\theta$ is either with respect to the Lebesgue measure for continuum angles or with respect to a discrete measure (a finite sum) for discrete angles.
 
The models presented have the following general model structure:
\begin{equation}
 \label{eq:genform}
     \partial_t f + v_0 \nabla \cdot \left( f  (1- \rho) \e(\theta) + a\phi \mathbf{p} f\right) = 
      D_T \nabla \cdot \left(   \mathcal{B}_1(\rho ) \nabla f + \mathcal{B}_2(f) \nabla \rho \right) + c \Dtheta   f.
\end{equation}
with $a \in \lbrace 0, 1 \rbrace$. In \eqref{eq:genform} the derivative operator $\nabla$ is the standard gradient with respect to the spatial variable $\x$, while the Laplacian $\Delta_\theta$ is either
\begin{itemize}
	\item the second derivative $\partial_{\theta \theta}f$ in the Brownian case,
	\item the second-order difference or discrete Laplacian 
	$${\cal D}^2f =  (f_{k+1} + f_{k-1} - 2f_k), $$
	with cyclic extension of the index $k$, 	in the case of fixed discrete rotations (or in one spatial dimension where there are only two possible orientations),
	\item the graph Laplacian with uniform weights
$${\cal D}_G f =  \sum_{j \neq k} (f_j - f_k),$$
in the run-and-tumble case \eqref{hybrid3} where arbitrary rotations are allowed.
\end{itemize}

Let us mention that similar structures and results hold true for graph Laplacians with other non-negative weights.
We provide an overview of the respective differential operators and constants for most of the presented models in Table \ref{tab:my_label}.

 \setlength{\tabcolsep}{12pt} 
\renewcommand{\arraystretch}{1.2}

 \begin{table}[ht]
     \centering
     \begin{tabular}{|c|c|c|c|c|c|c|}
          \hline
        Eq. Nr. & $\Delta_{\theta}$ &$a$ & $\mathcal{B}_1$ & $\mathcal{B}_2$ & $c$   \\
          \hline
          \eqref{model2}   & $\partial_{\theta \theta}$ &  $0 $ & 1 &  0 & $D_R$ \\
          \hline
            \eqref{model3} & $\partial_{\theta \theta}$ & $1$ &$(1- \rho)$ & $3  f$ & $D_R$ \\
            \hline 
            \eqref{lattice2D} & ${\cal D}^2$ &  $0$ & 1 & 0 & $\lambda$ \\
            \hline 
            \eqref{ASEP_2D} & ${\cal D}^2$ & $0$ & $(1-\rho)$ & $ f$ & $\lambda$\\
            \hline
            \eqref{model_hy} & $\partial_{\theta \theta}$ &  $0$& $(1- \rho)$& $ f$ & $D_R$ \\
            \hline 
            \eqref{model_wilson} &  ${\cal D}^2$ &  $1$& $(1- \rho)$ & $3  f$ &  $D_R$\\
            \hline 
            \eqref{hybrid3} &  ${\cal D}_G$ &  $1$& $(1- \rho)$ & $3  f$ &  $D_R$\\
            \hline
            \eqref{eq:MF_cross_diff_sys} & None &  $0$ & $(1- \rho)$ & $3 f$ & 0\\
            \hline
            \eqref{model3_passive} & None &  $0$ & $(1- \rho)$ & $ f$ & 0 \\
        \hline
     \end{tabular}
     \caption{Table recasting most models in the general form of \eqref{eq:genform}.}
     \label{tab:my_label}
 \end{table}
 
\paragraph{Small and large speed}
 
Natural scaling limits for the general system \eqref{eq:genform} are the ones for small and large speed, i.e., $v_0 \rightarrow 0$ and $v_0 \rightarrow \infty$, respectively. The first case is rather obvious, since at $v_0=0$ the model is purely diffusive, i.e.,
$$  \partial_t f   = 
      D_T \nabla \cdot (   \mathcal{B}_1(\rho ) \nabla f + \mathcal{B}_2(f) \nabla \rho) + c \Dtheta   f.$$
The model can then be written as a gradient-flow structure (or a generalised gradient structure in the case of discrete angles, see for example \cite{M2011,PRST2020}) for an entropy of the form 
 \begin{equation} \label{entropy_general}
{\cal E}(f) = \iint f \log f ~\ud \x ~\ud \theta +  
b_2 \int (1-\rho) \log (1 -\rho) ~\ud \x,
\end{equation}
with $b_2  \in \{0,1,3\}$ corresponding to the coefficients of ${\cal B}_2$.
In the case $v_0$ small but finite,  the gradient-flow structure is broken but we still expect the diffusive part to dominate. In particular, we expect long-time convergence to a unique stationary solution. 

In the case $v_0 \rightarrow \infty$ there are two relevant time scales. At a small time scale $L/{v_0}$, where $L$ is a typical length scale,  the evolution is governed by the first-order equation
$$\partial_\tau f +  \nabla \cdot \left( f  (1-\phi \rho) \e(\theta) + a\phi \mathbf{p} f \right) =  0,$$
where $\tau = t v_0/L$.
The divergence of the corresponding velocity field 
${\bf u} = (1-\phi \rho) \e(\theta) + a\phi \mathbf{p}$ is given by
$$ \nabla \cdot {\bf u} = - \phi \nabla \rho \cdot \e(\theta) + a \phi \nabla \cdot \mathbf{p}. $$
In particular in the case of $a=0$ we see that the question of expansion or compression of the velocity field is determined by the angle between $\nabla \rho$ and the unit vector $\e(\theta)$. Unless $\nabla \rho = 0$, the velocity field is compressible for a part of the directions and expansive for the opposite directions. A consequence to be expected is the appearance of patterns with almost piecewise constant densities (see, for example, Figures \ref{fig:model3_70_60} and \ref{fig:model4_70_60}). Inside the structures with constant densities ($\nabla \rho = 0$) the velocity field is incompressible, while the compression or expansion arises at the boundaries of such regions. This is rather described by a large time scale, i.e., the equation without time rescaling. Then one expects a slow interface motion, which is also observed in numerical simulations. In a simple case with only one direction this has been made precise in \cite{burger2008asymptotic}.
 
\paragraph{Small and large rotational diffusion}

The limit of small rotations rate $c \to 0$ corresponds to a more standard nonlinear Fokker-Planck system with a given linear potential,
$$     \partial_t f + v_0 \nabla \cdot \left( f  ((1-\phi \rho) \e(\theta) + a\phi \mathbf{p})\right) = 
      D_T \nabla \cdot (   \mathcal{B}_1(\rho ) \nabla f + \mathcal{B}_2(f) \nabla \rho), $$
as describe in Section \ref{s:nonact}. 
Models of this kind have been investigated previously, see for example \cite{Burger:2010gb, BHRW2016}. They tend to develop patterns such as jams or lanes, depending on the initial condition. This happens in particular for large speeds $v_0$ (see Figures \ref{fig:model4p_60_60} and \ref{fig:model4p_70_40}).

The case of large rotational diffusion $c\rightarrow \infty$ will formally lead to $f$ being constant with respect to $\theta$ at leading order. The corresponding equation at leading order can thus be obtained by averaging \eqref{eq:genform} in $\theta$. Since $f$ does not depend on $\theta$, the polarisation is zero, that is
$$ \mathbf{p} = \int_0^{2\pi} f   \e(\theta) ~ \ud \theta = 0, $$
and the transport term drops out in all the models. Indeed, the nonlinear diffusion terms in any case average to linear diffusion with respect to $\x$. Hence, the evolution of $f$ at leading order is governed by a linear diffusion equation.

\subsection{Wasserstein gradient flows}

We have seen above that microscopic models for externally activated particle have an underlying gradient-flow structure, which should ideally be maintained in the macroscopic limit. Adams et al. \cite{ADPZ2011} showed in their seminal work that then the Wasserstein metric arises naturally in the mean-field limit (under suitable scaling assumptions). However, this limit is only well understood in a few cases (for example for point particles) and rigorous results are often missing. In case of excluded-volume effects, as discussed in subsections \ref{sec:evi} and \ref{sec:hybrid}, the only known rigorous continuum models are derived in 1D \cite{Rost1984, BV2005, Gavish:2019tu}, with only approximate models for higher space dimension. 
We see that these approximate limits often lack a full gradient-flow structure, but are sufficiently close to it. In the following we give a brief overview on how Wasserstein gradient flows and energy dissipation provides useful a-priori estimates that can be used in existence proofs or when studying the long time behaviour of solutions. These techniques are particularly useful for systems with cross-diffusion terms, for which standard existence results do not necessarily hold.

We will outline the main ideas for functions $f = f(\x, \theta, t)$ where $\theta$ is either continuous or taking discrete values $\theta_k$ with $k = 1, \ldots m$. As before, we use $\xi = (\x, \theta)$. We say that a macroscopic model has a Wasserstein gradient-flow structure if it can be written as
 \begin{align}
 \label{e:w2sys}
    \partial_t f(\x, \theta, t)    &=
    \nabla_\xi \cdot \left( \mathcal{M}(f) \nabla_\xi
    w
    \right),
 \end{align}
 where $\mathcal{M}$ is the mobility operator and $w = \delta_f \mathcal{E}$ the variational derivative of an entropy/energy functional $\mathcal{E}$ with respect to $f$. Note that for discrete $\theta_k$, $k=1, \ldots m$ the mobility $\mathcal{M}$ is a positive definite matrix in $\R^{m \times m}$ and $\delta_f \mathcal{E}$ is replaced by the vector $\delta_{f_k} \mathcal{E}$. We have seen a possible candidate for energies in \eqref{entropy_general}; they usually comprise negative logarithmic entropy terms of the particle distribution and the total density (corresponding to linear and non-linear diffusion relating to the operators $\mathcal{B}_1$ and $\mathcal{B}_2$) as well as potentials.
 
If the system has a Wasserstein gradient-flow structure \eqref{e:w2sys} then the entropy $\mathcal{E}$ changes in time as
\begin{align}\label{e:entropydiss}
    \frac{d \mathcal{E}}{dt} = \iint \partial_t f  w \, \ud \x \ud \theta = -\iint \bar{\mathcal{M}}(w) \lvert \nabla_\xi w \rvert^2 \, \ud \x \ud \theta,
    \end{align}
where $\bar{\mathcal{M}}$ is the  mobility matrix $\mathcal M$ written in terms of the entropy variable $w$. If $\bar{\mathcal{M}}$ is positive definite, then the energy is dissipated. In the next subsection we will define an entropy for the general model \eqref{eq:genform} and show that the system is dissipative for several of the operator choices listed in Table \ref{tab:my_label}. 

Note that these entropy dissipation arguments are mostly restricted to unbounded domains and bounded domains with no-flux or Dirichlet boundary conditions. It is possible to generalise them in the case of non-equilibrium boundary conditions, as such discussed in Section \ref{s:bc}, but a general theory is not available yet. We will see in the next subsection that entropy dissipation may also hold for systems, which do not have a full gradient-flow structure. 
    
Since system \eqref{e:w2sys} is dissipative, we expect long time convergence to an equilibrium solution. The respective equilibrium solutions $f_\infty$ to \eqref{e:w2sys} then correspond to minimisers of the entropy $\mathcal{E}$. To show exponential convergence towards equilibrium it is often helpful to study the evolution of the so-called relative entropy, that is
\begin{align*}
    \mathcal{RE}(f, f_{\infty}) := \mathcal{E}(f) - \mathcal{E}(f_{\infty}) - \langle \mathcal{E}'(f_\infty),f-f_\infty\rangle.
\end{align*}
In general one wishes the establish so-called entropy-entropy dissipation inequalities for the relative entropy
\begin{align*}
    \frac{d\mathcal{RE}}{dt} \leq - C \mathcal{RE},
\end{align*}
with $C>0$. Then Gronwall's lemma gives desired exponential convergence. This approach is also known as the Bakry--Emery method, see \cite{BE1985}.

 We discussed the challenges in the rigorous derivation of continuum models in the previous sections and how often only formal or approximate limiting results are available. These approximate models are often 'close' to a full gradient flow, meaning that they only differ by higher order terms (which were neglected in the approximation). This closeness motivated the definition of so-called {\em asymptotic gradient flow}, see \cite{BBRW2017,bruna2017asymptotic}. A dynamical system of the form 
 \begin{align}\label{e:agf}
     \partial_t z = \mathcal{F}(z; \epsilon)
 \end{align}
 has a an asymptotic gradient-flow structure of order $k$ if 
 \begin{align*}
     \mathcal{F}(z; \epsilon) + \sum_{j=k+1}^{2k} \epsilon^j \mathcal{G}_j(z) = -\mathcal{M}(z; \epsilon) \mathcal{E}'(z, \epsilon),
 \end{align*}
 for some parametric energy functional $\mathcal{E}.$ For example, \eqref{eq:MF_cross_diff_sys} exhibits a GF structure if the red and blue particles have the same size and diffusivity, but lacks it for differently sized particles (a variation of the model not discussed here). The closeness of AGF to GF can be used to study for example its stationary solutions and the behaviour of solution close to equilibrium, see \cite{ARSW2020, ABC2018, BBRW2017}.
 
\subsection{Entropy dissipation}\label{sec:entropydiss}

Next we investigate the (approximate) dissipation of an appropriate energy for the general formulation \eqref{eq:genform}. The considered energy functional is motivated by the entropies of the scaling limits considered before. In particular we consider
 \begin{equation}
{\cal E}(f) = \iint f \log f + V(\x,\theta) f ~\ud \x~\ud \mu(\theta) + 
b_2 \int (1-\rho) \log (1 -\rho) ~\ud \x,
\end{equation}
for which the models can be formulated as gradient flows in the case $c=0$ (no active self-propulsion) with $b_2 \in \{0,1,3\}$ chosen  appropriately. For simplicity we set $\phi = 1$ as well as $D_T=1$ in the following. As before, we interpret integrals in $\theta$ with respect to the Lebesgue measure for continuum angles and with respect to the discrete measure (sum) in case of a finite number of directions. We recall that the potential $V$ is given by  
$$
V(\x, \theta) = - v_0 \, \e(\theta) \cdot \x = - v_0\, (\cos \theta x + \sin \theta y).
$$
In the following we provide a formal computation assuming sufficient regularity of all solutions. We have
\begin{align*}
    \frac{d \mathcal{E}}{dt} &= \iint \partial_t f  ( \log f + V - b_2 \log(1-\rho)) ~\ud \x~\ud \theta \\
    &= - \iint \nabla  \left [ \log f + V -  b_2 \log(1-\rho) \right]   
  \left\{-  v_0 f  [(1-\rho) \e(\theta) + a \mathbf{p} ] + 
         \mathcal{B}_1(f,\rho ) \nabla  f + \mathcal{B}_2(f,\rho) \nabla \rho \right \} 
      ~\ud \x~\ud \theta\\
    & \ \phantom{=} + c \iint    ( \log f + V - b_2\log(1-\rho))   \Delta_\theta f ~\ud \x~\ud \theta.
\end{align*}

Let us first investigate the last term.
Since $\rho$ is independent of $\theta$, using the properties of the generalised Laplacian $\Delta_\theta$ with periodic boundary conditions we have
$$ \int     \log(1-\rho) \Delta_\theta f~d\theta  = 
\log(1-\rho) \int      \Delta_\theta f~d\theta = 0.$$
Using the fact that $\Delta_\theta \e(\theta)$ is uniformly bounded in all cases, we find
\begin{align*}
    \iint    [ \log f + V -  b_2 \log(1-\rho) ]   \Delta_\theta f ~\ud\x~\ud\theta &= 
    - \iint {\cal F}_\theta(f) - v_0 \Delta_\theta \e(\theta) \cdot \x f ~d\x~d\theta, \\
    &\leq C |v_0| \int |\x| f~\ud\x~\ud\theta = C |v_0| \int |\x| \rho~\ud\x,
\end{align*}
where ${\cal F}_\theta(f) \geq 0$ is the Fisher information with respect to the  generalised Laplacian $\Delta_\theta$ 
$$ {\cal F}_\theta(f) = 
\begin{cases}
	\displaystyle \frac{|\partial_\theta f|^2}f & \text{for } \Delta_\theta = \partial_{\theta \theta}, \\
 \displaystyle \frac{|f_{k+1}-f_k|^2}{M(f_k,f_{k+1})}& \text{for } \Delta_\theta = {\cal D}^2, \\
\displaystyle \sum_j \frac{|f_{j}-f_k|^2}{M(f_j,f_k)} \qquad  & \text{for } \Delta_\theta = {\cal D}_G,
\end{cases}
$$
where 
$$M(f,g) = \frac{f-g}{\log(f) - \log(g)}$$
is the logarithmic mean. 

Now we further investigate the first term for the models with $a =0$ (no $\p$ term in the equation for $f$), where, for the respective $b_2 $ we obtain
\begin{multline*}
\iint \nabla  [ \log f + V - b_2 \log(1-\rho)]  \left[
   v_0    f  (1- \rho) \e(\theta) -    \mathcal{B}_1(f,\rho ) \nabla f - \mathcal{B}_2(f,\rho) \nabla \rho   \right ]~\ud\x~\ud\theta  \\
 = - \iint  f(1-\rho)  |\nabla  [ \log f + V - b_2\log(1-\rho)] |^2~\ud\x~\ud\theta  \leq 0.
\end{multline*}
Overall we finally find
$$ 
\frac{d \mathcal{E}}{dt} \leq C ~|v_0|~\int |\x| \rho~\ud\x \leq  C ~|v_0|~ \sqrt{\int |\x|^2 \rho~\ud\x} . 
$$
Thus, the growth of the entropy in time is limited by the second moment. Note that for $a=1$ one can employ analogous reasoning to obtain the above negative term. However it is unclear how to control the  additional term 
$\iint \nabla  [ \log f + V - c\log(1-\rho)]  
   v_0   {\bf p} f ~\ud\x~\ud\theta$. The bounds obtained provide useful a-priori estimates, which can be used in existence results and to study the long-time behaviour, see for example \cite{BSW2012, J2015}.


\section{Boundary effects}\label{s:bc}

So far, we have focused on domains with periodic boundary conditions. In this section we discuss non-zero flux boundary conditions, which can be used to impose non-zero steady currents and externally-activate or force  
 the passive models described in Section \ref{s:nonact} out of equilibrium \cite{Cates:2014tr}. 
We remark that the in-flux boundary conditions are difficult to deal with in the case of interacting continuous random walks. Thus we only mention a few aspects and comment in more detail on the time-discrete situation which is easier to tackle, see remark \ref{rem:time_discrete}.

\subsection{Mass conserving boundary conditions} We first discuss conditions (other than periodic boundaries) that conserve the total mass, i.e., the total number of particles in the microscopic models, or the integral of the density $\phi$ in the macroscopic models.
In case of the coupled SDE model \eqref{sde_model}, we are interested in conditions that ensure that particles remain inside the domain. Intuitively, particles need to be reflected whenever they hit the boundary. However, as we are dealing with a problem that is continuous in time, we have to ensure that the particle path remains continuous. In his seminal paper \cite{Skorokhod1961_bounded} Skorokhod solved this problem by introducing an additional process that increases whenever the original process hits the boundary, see \cite{Pilipenko2014_reflection} for a detailed discussion. 
For the microscopic models on a lattice, such boundary conditions 
correspond to aborting any jumps that would lead a particle outside of the domain. 
For the macroscopic models, mass conservation corresponds to no-flux boundary conditions that are implemented by setting the normal flux over the boundary to zero, i.e.,
\begin{align}\label{eq:noflux}
\J \cdot \n = 0 \text{ a.e. in } \Upsilon \times (0,T),
\end{align}
where, using the general form \eqref{eq:genform}, the flux density is given as 
\begin{align}\label{eq:J_general}
    \J =v_0  ( f  ((1-\phi \rho) \e(\theta) + a\phi \mathbf{p})) -      D_T (  \mathcal{B}_1(\rho ) \nabla f + \mathcal{B}_2(f) \nabla \rho).
\end{align}

\subsection{Flux boundary conditions}
Apart from periodic or no-flux boundary conditions, there is also the possibility for boundary conditions that allow for the in- or outflow of particles (mass) via the boundary. Such effects are of particular interest in the context of this chapter, since they yield an out-of-equilibrium system even if the motion of the particles within the domain is purely passive (i.e., due to diffusion). 

For the SDE model \eqref{sde_model}, such boundary conditions correspond to partially reflecting or radiation conditions.
Intuitively, once a particle reaches the boundary it is, with a certain probability, either removed or otherwise reflected, see \cite{grebenkov2006partially} and \cite[Section 4]{Lions1984}.
For the discrete models of Section \ref{sec:discrete}, let us consider first the special case of a single species in two dimensions with two open and to closed boundaries. This corresponds to the asymmetric simple exclusion process (ASEP) with open boundary conditions, the paradigmatic models in non-equilibrium thermodynamics, \cite{chou2011_nonequilibrium}. 
The dynamics of such a process is well understood and can be solved explicitly  \cite{Derrida1993:TASEP,derrida1998_exactly} (see also \cite{Wood2009:TASEP_boundary}). 
We denote by $\alpha$ and $\beta$ the rates by which particles enter (at the left boundary) and exit (at the right boundary) the lattice. Then, the  key observation here is that in the steady state, system can be in one of three distinct states, characterised by the value of the one-dimensional current and the density as follows
\begin{itemize}
    \item \emph{Low density} or \emph{influx limited} ($\alpha < \min\{\beta, 1/2\}$): the density takes the value $\alpha$ and the flux $\alpha(1-\alpha)$.
    \item \emph{High density} or \emph{outflux limited} ($\beta < \min\{\alpha, 1/2\}$): the density is $1-\beta$ and the flux $\beta(1-\beta)$.
    \item \emph{Maximal current} ($\alpha , \beta > 1/2$): the density is $1/2$ and the flux $1/4$.
    \end{itemize}
A similar behaviour can be verified for the macroscopic passive model \eqref{eq:MF_cross_diff} (or also \eqref{model_lattice1D_number_densities} with $\lambda =0$) for a single species on the domain $\Omega = [0,L]$, which reduces to a single equation for the unknown density $r$, i.e.,
\begin{align*}
    \partial_t r + \partial_x j = 0 \text{ with } j = -D_T \partial_x r + r(1-r) \partial_x V.
\end{align*}
We supplement the equation with the flux boundary conditions 
\begin{align}\label{eq:inoutflux}
    -j \cdot n = \alpha (1- r) \text{ at } x = 0 \text{ and } j \cdot n = \beta r \text{ at } x = 1,
\end{align}
see \cite{Burger2016}. Indeed, one can show that for positive $D_T> 0$, stationary solutions are close to one of the regimes and as $D_T \to 0$, these attain the exact values for flux and density. Interestingly, for positive $D_T$ it is possible to enter the maximal current regime for values of $\alpha$ and $\beta$ strictly less than $1/2$. The long time behaviour of these equations, using entropy--entropy-dissipation inequalities, has been studied in \cite{Burger2016}.

For the macroscopic active models \eqref{model3} and \eqref{model_hy}, a similar condition can be formulated for the unknown quantity $f$. However, as $f$ depends not only on $\bf{x}$ and $t$ but also on the angle $\theta$, the coefficients may also depend on it. In the most general situation we obtain
\begin{align}\label{eq:flux_bc}
  \J \cdot \n =-  \alpha(\theta, \n)  {(1- \phi\rho)} +  \beta(\theta, \n)f,
\end{align}
with $\J$ defined in \eqref{eq:J_general}.
Here, the choice of the functions $\alpha$ and $\beta$ is subject to modelling assumptions or properties of microscopic stochastic models for the in- and outflow. Typically one has a separation into inflow- and outflow regions, which means that $\alpha$ is supported on inward pointing directions $\e(\theta) \cdot \n > 0$, while $\beta$ is supported outward pointing directions $\e(\theta) \cdot \n > 0$. 

\subsection{Other boundary conditions}
Let us also discuss other types of boundary conditions. Homogeneous Dirichlet boundary conditions can be applied to all types of models: for the SDE \eqref{sde_model}, one has to remove a particle once it reaches the boundary. The same holds for the discrete random walk models. For the macroscopic models, one sets the trace at the boundary to zero. Finally, also mixed boundary conditions are possible, combining the effects described above on different parts of the boundary.
Another type of boundary condition useful in the context of self-propelled organisms are no-slip or alignment type boundary conditions, whereby the particles align their orientations with the boundary ($\e(\theta) \cdot \n = 0$). A notable example of this can be seen in ant foraging networks and lab experiments with ants walking on bridges \cite{10.1098/rspb.2004.2990,Dobramysl:2021wy}.

\begin{remark}[Boundary conditions for discrete time random walks] \label{rem:time_discrete}
We briefly comment on the situation for time-discrete random walks, that is when the SDE \eqref{sde_model} is replaced by the time-discrete system
\begin{subequations}
	\label{sde_model_discrete}
\begin{align}
\label{sde_x_discrete}
	\X_i(t+\Delta t) &= \X_i(t) + \Delta t\sqrt{2 D_T} \zeta_i - \Delta t \nabla_{\x_i} U + \Delta t v_0 \e(\Theta_i),\\
	\label{sde_angle_discrete}
	\Theta_i(t + \Delta t) &= \Theta_i(t) + \Delta t\sqrt{2 D_R} \bar \zeta_i - \Delta t\partial_{\theta_i} U,
\end{align}
\end{subequations}
for some time step size $\Delta t > 0$ and where $\zeta_i, \, \bar \zeta_i$ are normally distributed random variables with zero mean and unit variance. To implement boundary conditions, one has to calculate the probability that ${\bf{X}}_i(t + \Delta t) \notin \Omega$ (considering also the case that the particles leaves the domain but moves back into it within the time interval $[t, t+\Delta t]$), see \cite{Andrews2004_time_discrete} for detailed calculations in the case of pure diffusion. If a particle is found to have left the domain, it can either be removed with probability one (corresponding to homogeneous Dirichlet boundary conditions) or less than one, called a partially reflective boundary condition (corresponding to Robin boundary conditions). In our setting, this probability can depend on the current angle of the particle, $\Theta_i(t)$, allowing for additional modelling. 
It is also possible to add a reservoir of particles at the boundary to implement flux boundary conditions in the spirit of \eqref{eq:flux_bc} by prescribing a probability to enter the domain. In the case of excluded volume, the probability to enter will depend on the number of particles close to the entrance.
\end{remark}

\section{Active crowds in the life and social science}\label{sec:applications}

\subsection{Pedestrian dynamics} \label{sec:pedestrian}

A prominent example of active and externally activated dynamics in the context of socio-economic applications is the motion of large pedestrian crowds. There is an extensive literature on mathematical modelling for pedestrians in the physics and the transportation community, which is beyond the scope of this paper. We will therefore review the relevant models in the context of active crowds only and refer for a more comprehensive overview to \cite{CPT2014, MF2018}. 

\paragraph{Microscopic models for pedestrian flows}
Microscopic off-lattice models are the most popular approach in the engineering and transportation research literature. Most software packages and simulations are based on the so called social force model by Helbing  \cite{Helbing1995:social,Helbing2000:social}. The social force model is a second order SDE model, which does not take the form of active models considered here. However, it is easy to formulate models for pedestrians in the context of active particles satisfying \eqref{sde_model}. For example, assume that all pedestrians move with the same constant speed in a desired direction $\Theta_d$  avoiding collisions with others. Then their dynamics can be described by the following second order system:
\begin{subequations}    
\label{e:activepedestrians}
\begin{align}
    \ud \X_i &= -\nabla_{\X_i} U \ud t  + v_0 \frac{e(\Theta_i)- \Theta_d}{\tau}dt + \sqrt{2 D_T}\, \ud \W_i\\
    \ud \Theta_i &= -\partial_{\Theta_i} U \ud t + \sqrt{2D_R}\,\ud\W_i.
\end{align}
\end{subequations}
The potential $U$ takes the form \eqref{e:U}, where the pairwise interactions $u$ should be related to the likelihood of a collision. One could for example consider
\begin{align*}
    u(\lvert \X_i - \X_j \rvert/\ell, \Theta_i - \Theta_j) = C \frac{\Theta_i - \Theta_j}{\lvert \X_i - \X_j\rvert},
\end{align*}
where $C \in \mathbb{R}^+$ and $\ell$ relates to the personal comfort zone. Another possibility corresponds to a Lennard Jones type potential to model short range repulsion and long range attraction. 
Another popular microscopic approach are so-called cellular automata, which correspond to the discrete active and externally activated models discussed before. In cellular automata a certain number of pedestrians can occupy discrete lattice sites and individuals move to available (not fully occupied) neighbouring sites, with transition rates. These transition rates may depend on given potentials, as discussed in the previous sections, which relate to the preferred direction.

There is also a large class of microscopic on-lattice models, so called cellular automata, see \cite{Kirchner2002:Cellular}, which relate to the microscopic discussed in Section \ref{sec:discrete_passive}. In cellular automata pedestrians move to neighbouring sites at given rates, if these sites are not already occupied. Their rates often depend on an external given potential, which relates to the desired direction $\Theta_d$. Cellular automata often serve as the basis for the macroscopic pedestrian models, which will be discussed in the next paragraph, see for example \cite{BMP2011, Burger2016:sidestepping}. 

\paragraph{Macroscopic models for pedestrian flows}
Mean field models derived from microscopic off-lattice approaches have been used successfully to analyse the formation of directional lanes or aggregates in bi-directional pedestrian flows. This segregation behaviour has been observed in many experimental and real-life situations. Several models, which fall into the category of externally activated particles introduced in Section \ref{sec:discrete_passive}, were proposed and investigated in this context. These models take the form \eqref{eq:MF_cross_diff}, in which the densities $r$ and $b$ relate to different directions of motion. For example in the case of bi-directional flows in a straight corridor 'red particles' correspond to individuals moving to the right, while blue ones move to the left. We will see in Subsection \ref{sec:numerics1} that we can observe temporal as well as stationary segregated states. Depending on the initial and inflow conditions directional lanes or jams occur. Then the gradient-flow structure can then be used to investigate the stability of stationary states using for example the respective entropy functionals. Due to the segregated structure of stationary solutions, one can also use linear stability analysis around constant steady states to understand for example the formation of lanes, see \cite{Pietschmann2011:lane}.

More pronounced segregated states and lanes can be observed when allowing for side-stepping. In the respective microscopic on lattice models, individuals step aside when approached by a member of the other species. The respective formally derived mean-field model has a perturbed gradient-flow structure, which can be used to show existence of solutions, see \cite{Burger2016:sidestepping}. 
More recently, a model containing both and active and a passive species has been studied in \cite{Cirillo2020:ActivePassive}.

\subsection{Transport in biological systems}\label{sec:biological_transport}
Another example where active particles play an important role are transport process in biological systems. We will discuss two important types of such processes in the following - chemotaxis and transport in neurones. 

\subsubsection*{Chemotaxis}

We consider bacteria in a given domain that aim to move along the gradient of a given chemical substance, called chemo-attractant and modelled by a function $c:\Omega \to  \R_+$. Due to their size, bacteria cannot sense a gradient by, say, comparing the value of $c$ at their head with that at their tail. 
Thus, they use a different mechanism based on comparing values of $c$ at different time instances and different points in space, called run-and-tumble. In a first step, they perform a directed motion into a fixed direction (run), then rotate randomly (tumble). These two steps are repeated, however, the probability of tumbling depends on $c$ as follows: If the value of $c$ is decreasing in time, bacteria tumble more frequently as they are not moving up the gradient. If the value of $c$ increased, they turn less often. 
Roughly speaking, this mechanism reduces the amount of diffusion depending on the gradient of $c$. Here, we consider a slightly different idea that fits into the hybrid random walk model introduced in \eqref{hybrid2}, assuming $D_T$ to be small (run) and the rate of change for the angle depends on $c$. To this end $\lambda$ is taken different for each angle (thus denoted by $\lambda_k$) and is assumed to depend on the difference of the external signal $c$ at the current and past positions, only. Denoting by $t_k$, $k=1,2, \ldots$ the times at which the angle changes, at time $t_n$ we  have $\lambda_k  = \lambda_k((c(\mathbf{X}_i(t_n))-c(\mathbf{X}_i(t_{n-1}))$. 
Additionally, we introduce a fixed base-line turning frequency $\bar\lambda$, and consider
$$
\lambda_k = \bar \lambda +  (c(\mathbf{X}_k(t_{n-1})) - c(\mathbf{X}_k(t_n))),
$$ 
Now going from discrete to time-continuous jumps, i.e., $t_{n} - t_{n-1} \to 0$, and appropriate rescaling, we obtain via the chain rule
$$
\lambda_k = \bar \lambda -  \dot{\mathbf{X}}_k\cdot \nabla c(\mathbf{X}_i).
$$
However, due to the stochastic nature of the equation governing the evolution $\mathbf{X}_k$, its time derivative is not defined. Thus, as a modelling choice, we replace this velocity vector by $v_0 \e(\theta_k)$, i.e., the direction of the active motion of the respective particle. This is also motivated by the fact that for $D_T=0$ and $U=0$ in \eqref{hybrid_x}, this is exact. We obtain
$$
\lambda_k = \bar \lambda -  v_0\e(\theta_k)\cdot \nabla c(\mathbf{X}_i).
$$
In the particular case on one spatial dimension with only two possible angles (denoted by $+$ and $-$) and for $v_0=1$ this reduces to
$$
\lambda_\pm = \bar\lambda \mp \partial_x c,
$$
which is exactly the model analysed in \cite{Ralph:2020cj}. There, it was also shown that using an appropriate parabolic scaling, one can obtain a Chemotaxis-like model with linear transport but non-linear diffusion in the diffusive limit.

\subsubsection*{Transport in neurones}
Another interesting example are transport processes within cells and we focus on the example of vesicles in neurones. Vesicles are small bubbles of cell membrane that are produced in the cell body (soma) and are then transported along extensions of the cell called axons. The transport itself is carried out by motor proteins that move along microtubules and are allowed to change their direction of motion. 
This situation can be modelled using the discrete random walks from Section \ref{sec:discrete} by considering the one-dimensional case which, in the macroscopic limit, yields equations \eqref{model_lattice1D}. 
Since we are now dealing with two species $f_-$ and $f_+$, denoting left- and right-moving complexes, we also have to adopt our boundary conditions as follows: Denoting by $j_+$ and $j_-$ the respective fluxes,
\begin{align*}
    -j_+ &= \alpha_+ (1-\phi\rho),  \quad j_- = \beta_- f_- \qquad\text{ at } x = 0,\\
    -j_- &= \alpha_- (1-\phi\rho),  \quad j_+ = \beta_+ f_+ \qquad\text{ at } x = 1.
\end{align*}
System \eqref{model_lattice1D} has, to the best of our knowledge, not yet been considered with these boundary conditions. From an application point of view, it is relevant to study whether these models are able to reproduce the almost uniform distribution of motor complexes observed in experiments, see \cite{Bressloff2015_democracy, Bressloff2016_exclusion} for an analysis.

More recently, the influence of transport in developing neurites has been studied in \cite{humpert_role_2019} with an emphasis on the mechanism that decides which of the growing neurites becomes an axon. To model this situation, the concentration of vesicles at some and growth cones is modelled separately by ordinary differential equations which are connected to to instances of \eqref{eq:MF_cross_diff} via flux boundary conditions. 

\section{Numerical simulations}\label{sec:numerics}

\subsection{One spatial dimension} \label{sec:numerics1}

In the following, we present numerical examples in one spatial dimension comparing a subset of models presented above. All simulations in this subsection are based on a finite element discretisation in space (using P$1$ elements). The time discretisation is based on the following implicit-explicit (IMEX) scheme 
\begin{equation*}
      \frac{f^{n+1} - f^n}{\tau} + v_0 \nabla \cdot [ f (1-\phi \rho^n) \e(\theta) + a\phi \mathbf{p} f] = 
      D_T \nabla \cdot (   \mathcal{B}_1(\rho^n ) \nabla f^{n+1} + \mathcal{B}_2(f^n) \nabla \rho^{n+1}) + c\Dtheta   f^n,
\end{equation*}
in which the superscript index $n$ refers to the $n$th time step, that is $t^n = n \tau$, $\tau>0.$
Here transport and rotational diffusion are taken explicitly, while in the diffusive part terms of second order are treated implicitly. Thus, in every time step, a linear system has to be solved. All schemes were implemented using the finite element library NgSolve, see \cite{Schöberl1997}. 

We will illustrate the behaviour of solutions for  models \eqref{model_lattice1D_number_densities}, \eqref{e:aa_cross_sys}, \eqref{eq:MF_cross_diff}, in case of in- and outflux \eqref{eq:inoutflux}, no-flux \eqref{eq:noflux} or periodic boundary conditions in case of two species, referred to as red $r$ and blue $b$ particles. We use subscript $r$ and $b$, when referring to their respective in- and outflow rates as well as diffusion coefficients. Note that while for the models \eqref{model_lattice1D_number_densities}, \eqref{eq:MF_cross_diff}, the one-dimensional setting is meaningful, for model \eqref{e:aa_cross_sys}, the simulations are to be understood as two-dimensional but with a potential that is constant in the second dimension.
For all simulations, we discretised the unit interval into $150$ elements and chose time steps of size $\tau = 0.01$.
\subsubsection*{Flux boundary conditions} 
\begin{figure}
    \centering
    \subfigure[$t=0.5$]{
    \includegraphics[width=0.23\textwidth]{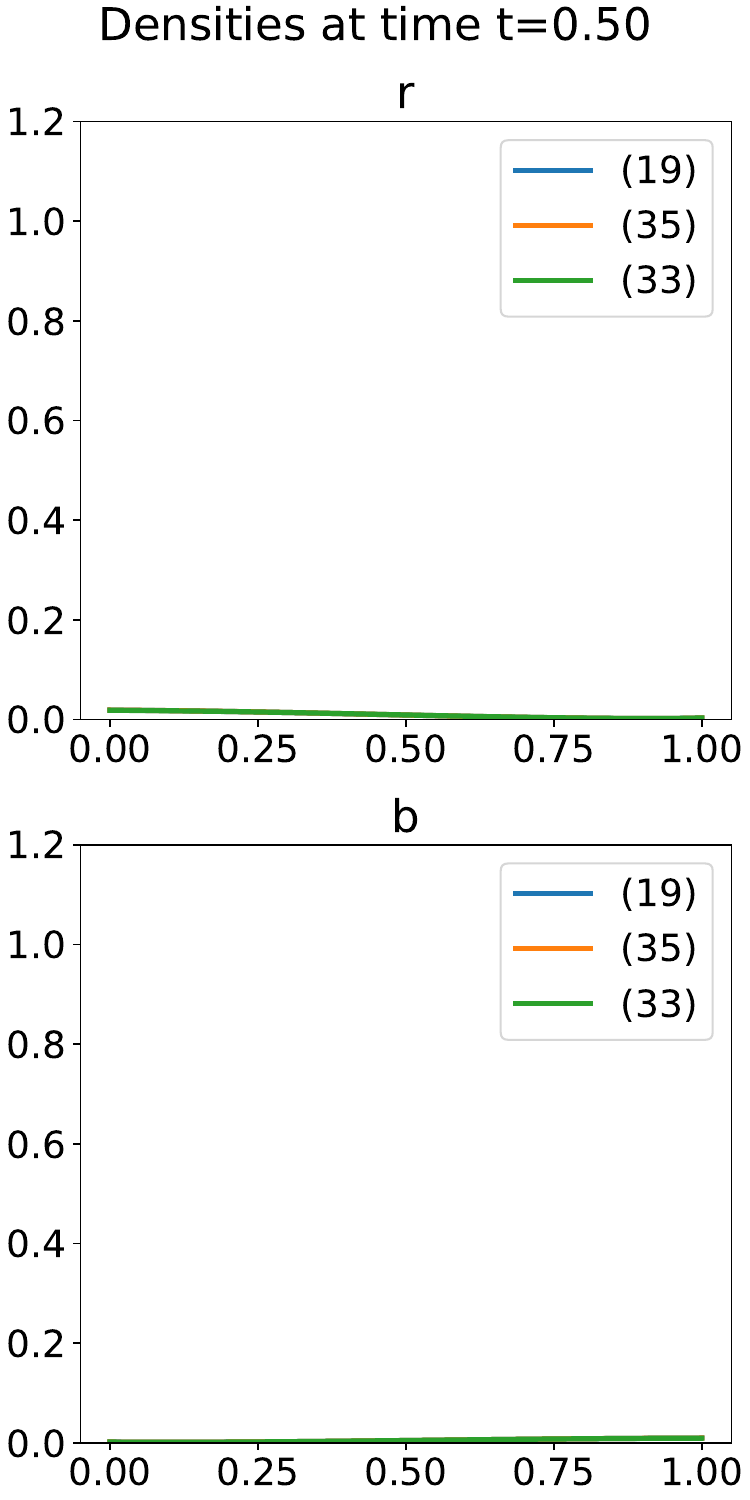}
    }
    \subfigure[$t=2$]{
    \includegraphics[width=0.23\textwidth]{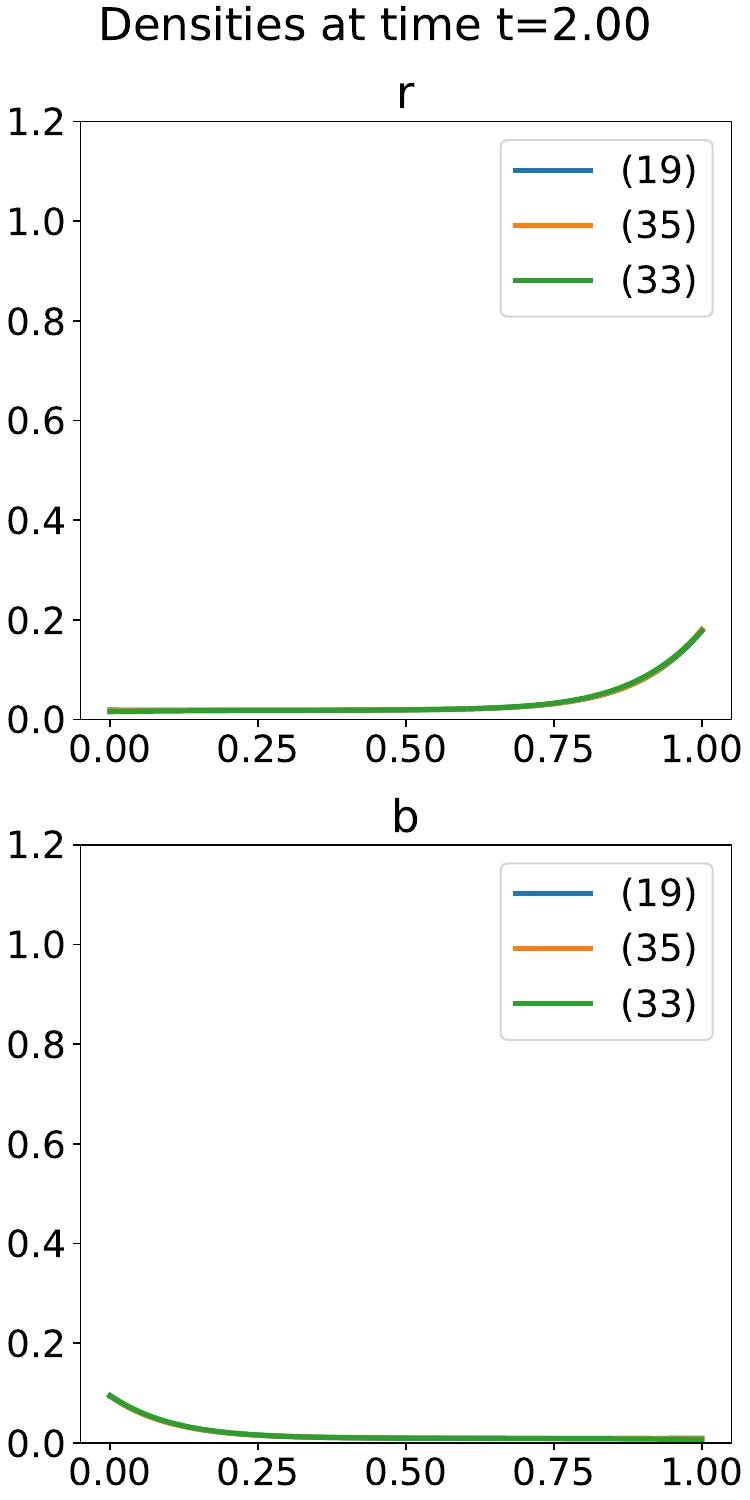}
    }
    \subfigure[$t=3$]{
    \includegraphics[width=0.23\textwidth]{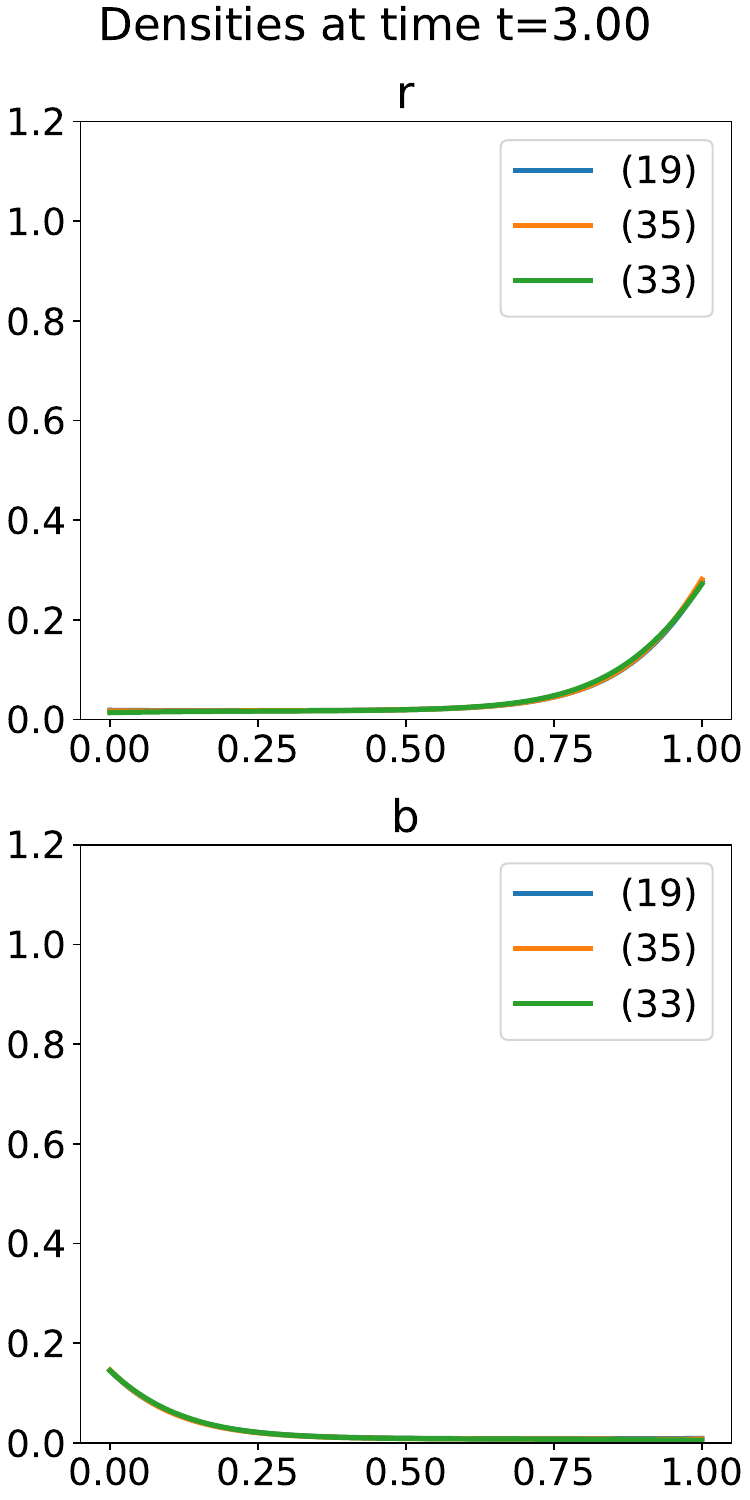}
    }
    \subfigure[$t=200$]{
    \includegraphics[width=0.23\textwidth]{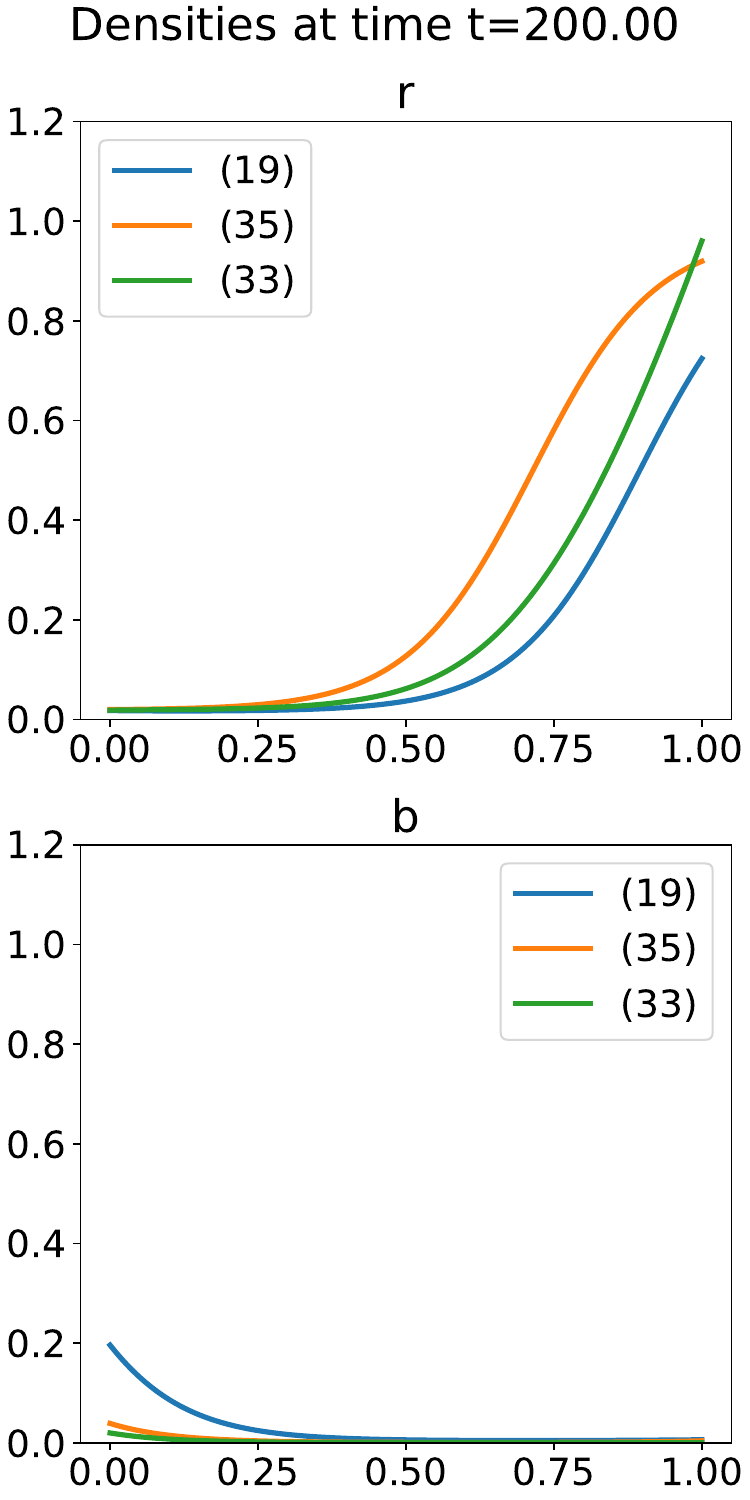}
    }
    \caption{Flux boundary conditions with: $\lambda = 0.01$, $D_r = 0.1,\,  D_b = 0.1,\, \alpha_r = 0.02, \, \beta_r = 0.01, \, \alpha_b = 0.01, \, \beta_b = 0.02$ which yields the influx-limited phase for species $r$ and outflux-limited for $b$.}
    \label{fig:Regime2}
\end{figure}

\begin{figure}
    \centering
    \subfigure[$t=0.5$]{
    \includegraphics[width=0.23\textwidth]{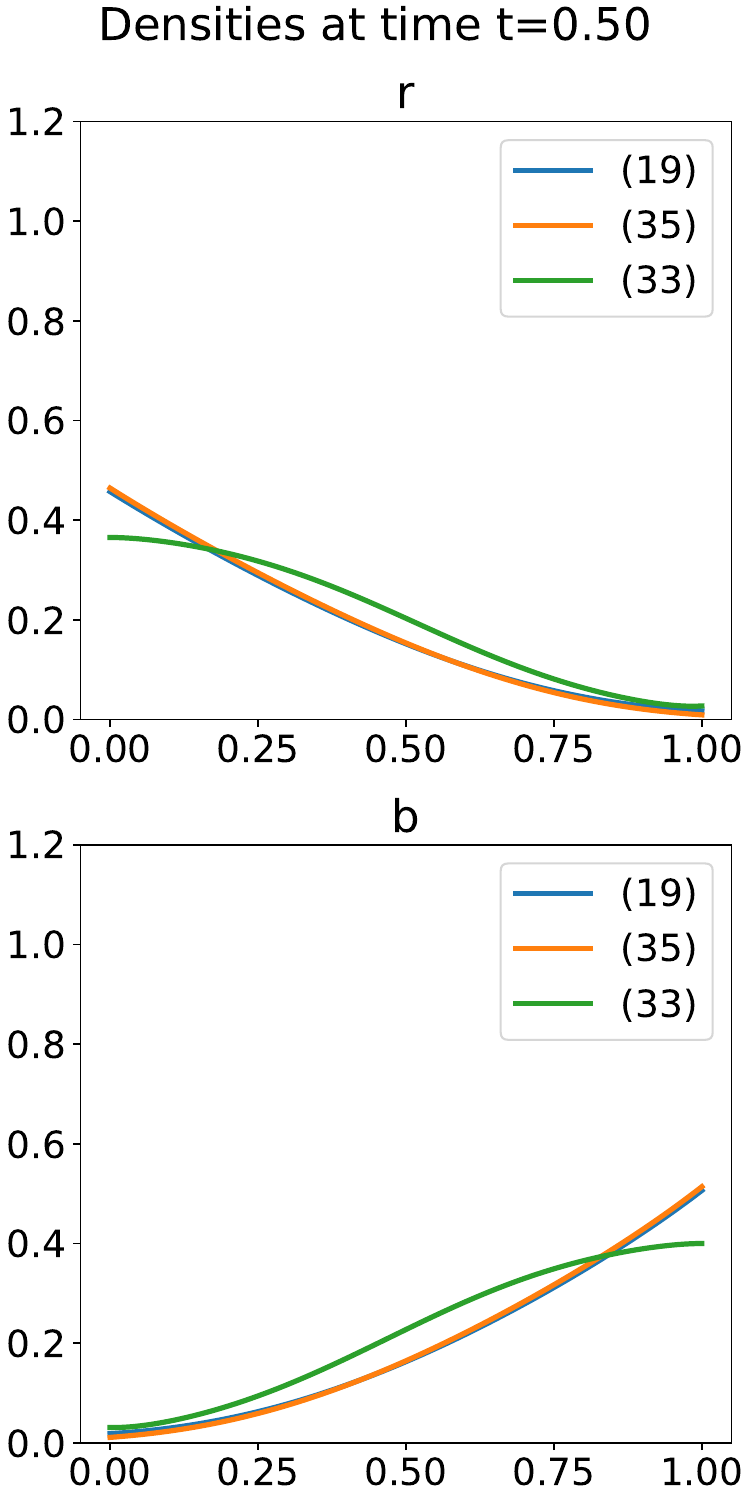}
    }
    \subfigure[$t=2$]{
    \includegraphics[width=0.23\textwidth]{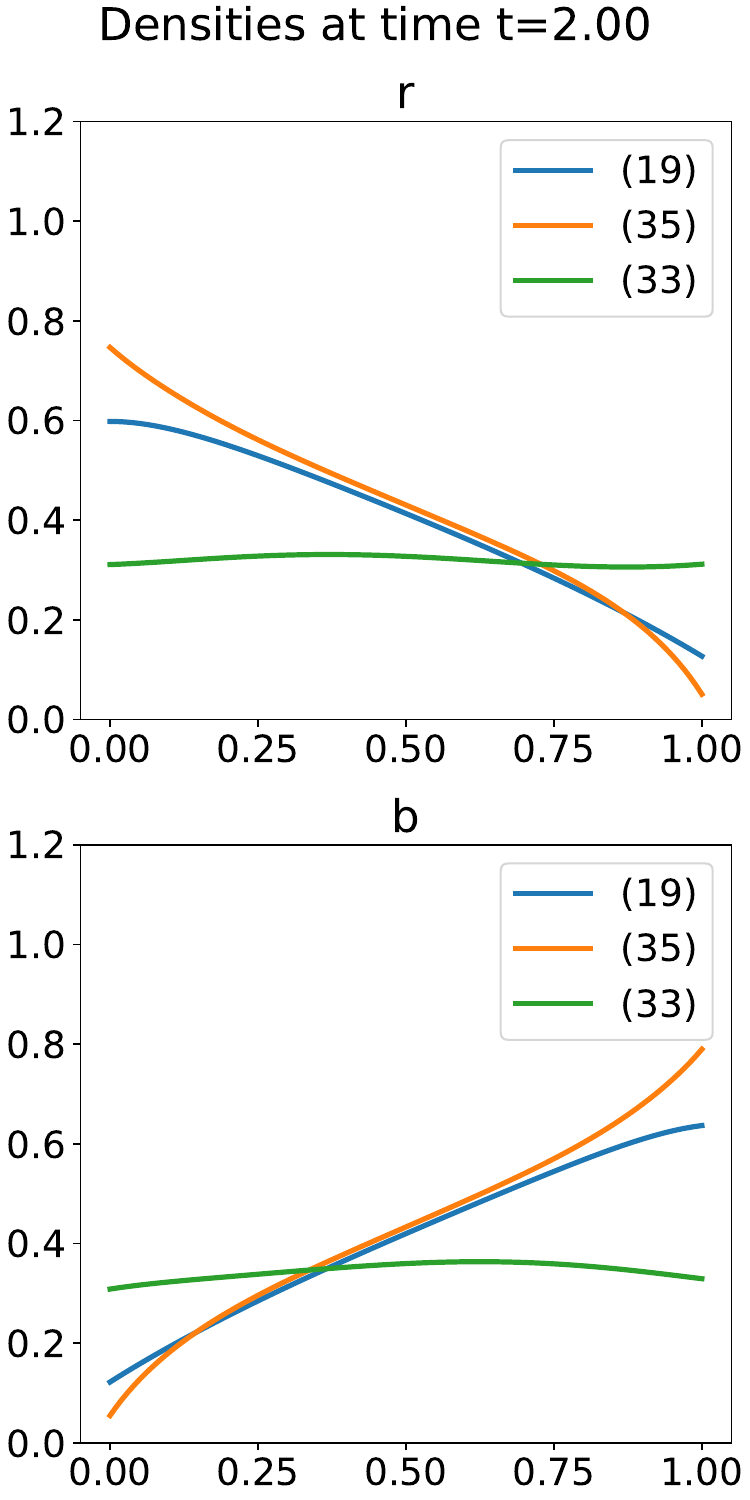}
    }
    \subfigure[$t=3$]{
    \includegraphics[width=0.23\textwidth]{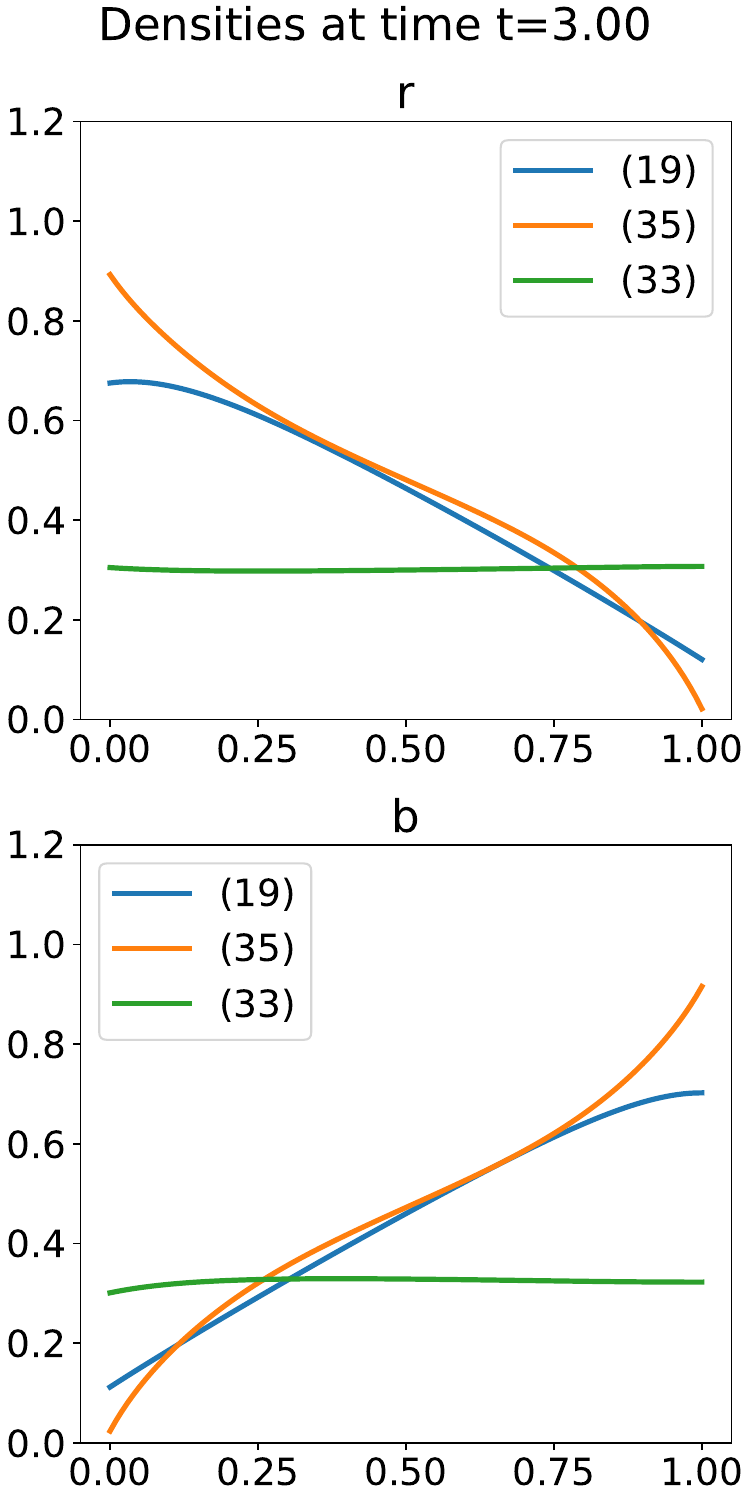}
    }
    \subfigure[$t=200$]{
    \includegraphics[width=0.23\textwidth]{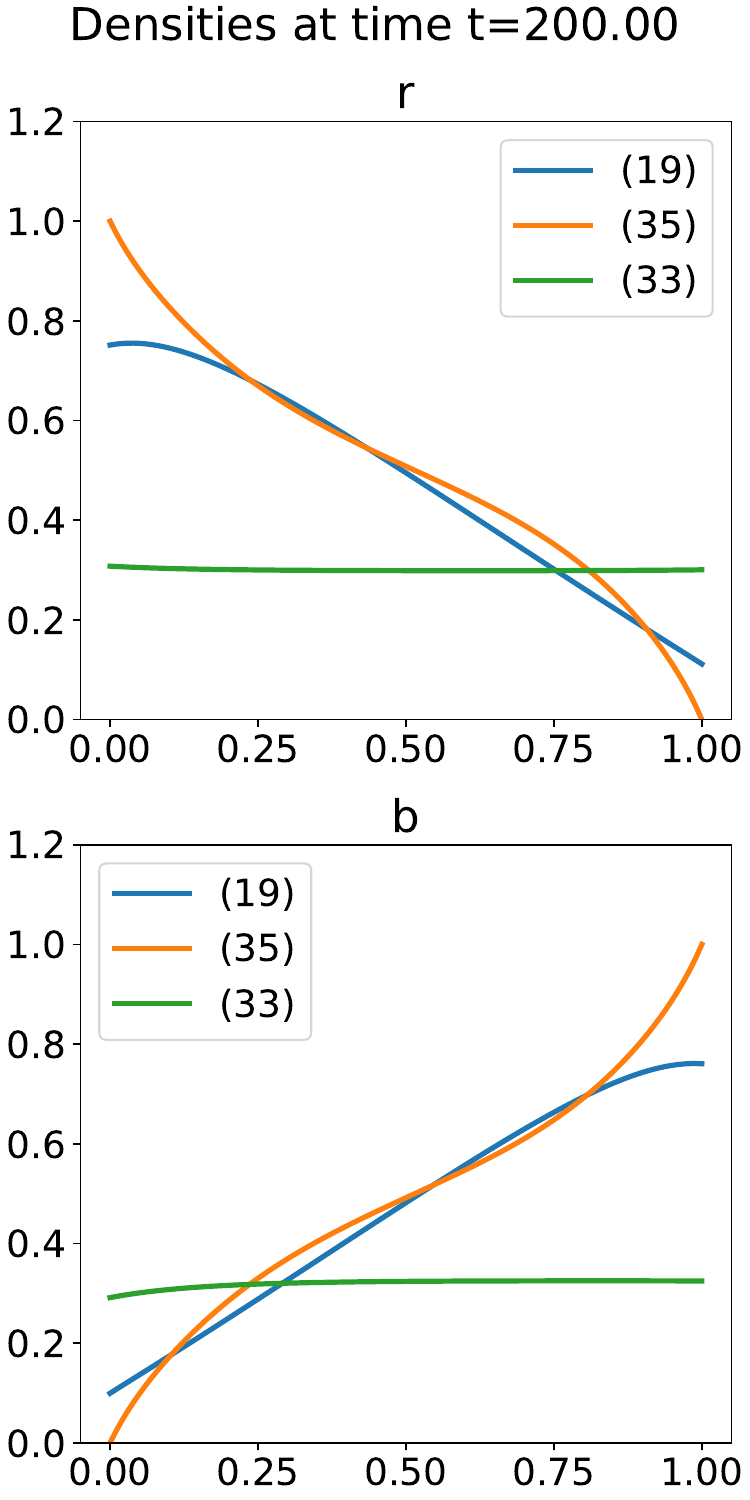}
    }
    \caption{Flux boundary conditions with: $\lambda = 0.01$, $D_r = 0.1,\,  D_b = 0.1,\, \alpha_r = 0.6, \, \beta_r = 0.8, \, \alpha_b = 0.7, \, \beta_b = 0.9$ which yields the maximal current phase.}
    \label{fig:Regime3}
\end{figure}

Figures \ref{fig:Regime2} and \ref{fig:Regime3} show density profiles for the respective models at time $t=0.5,\, 2,\, 3, \, 30$. In Figure \ref{fig:Regime2}, we chose rather low rates (in particular below $1/2$) and with $\alpha_r > \beta_r$ as well as $\alpha_b < \beta_b$ which resulted in species $r$ being in a outflux limited and species $b$ an influx limited phase. We observe that for these low rates, all models are quite close to one another, yet with different shapes of the boundary layers. Model \eqref{model_lattice1D_number_densities}, having a linear diffusion term, showing a different slope as \eqref{eq:MF_cross_diff} where cross-diffusion seems to play a role an \eqref{e:aa_cross_sys} being in between.

\begin{figure}
    \centering
    \subfigure[$\alpha_r=0.02$, $\beta_r=0.01$, $\alpha_b=0.01$,  $\beta_b=0.02$]{
    \includegraphics[width=0.275\textwidth]{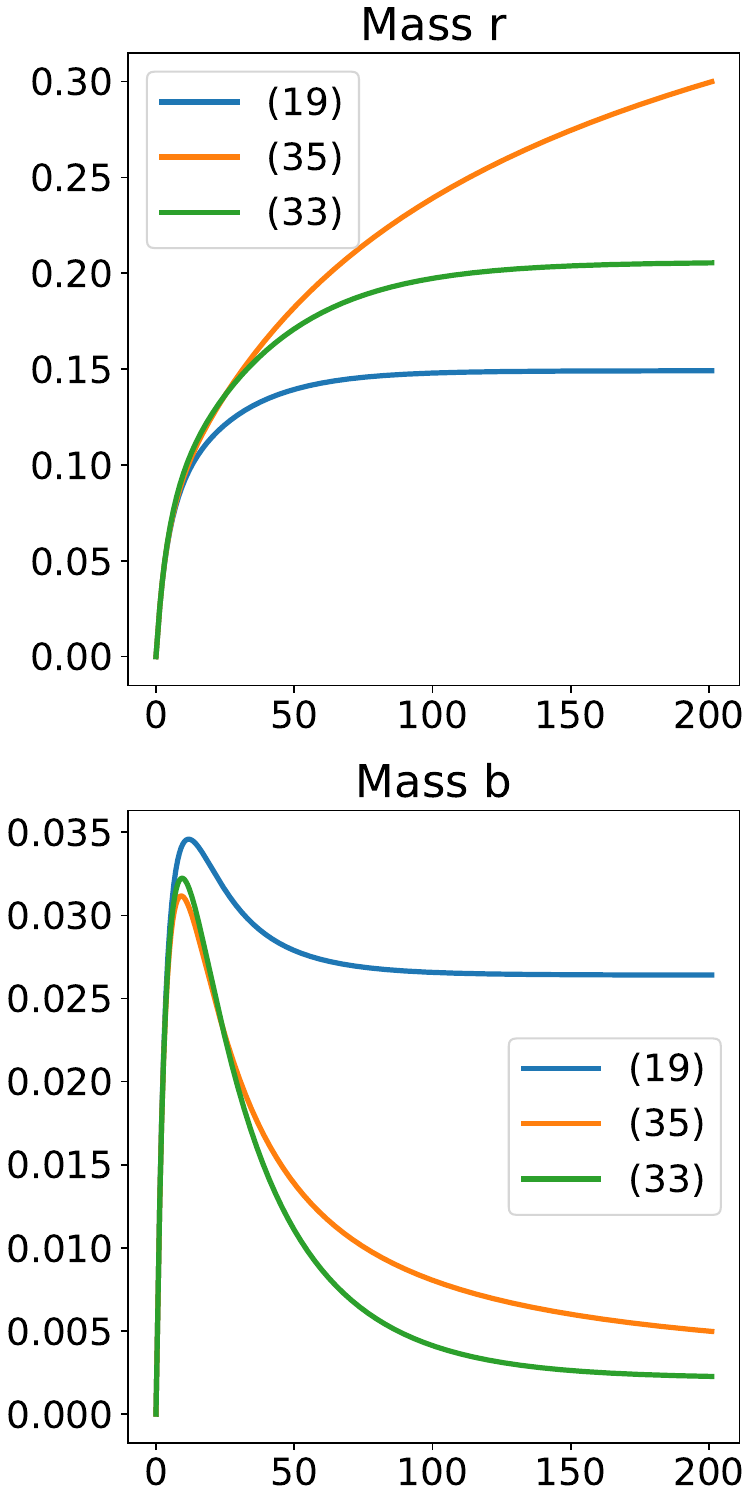}
    }
    \hspace{0.2cm}
    \subfigure[$\alpha_r=0.6$, $\beta_r=0.8$,   $\alpha_b=0.7$,  $\beta_b=0.9$]{
    \includegraphics[width=0.275\textwidth]{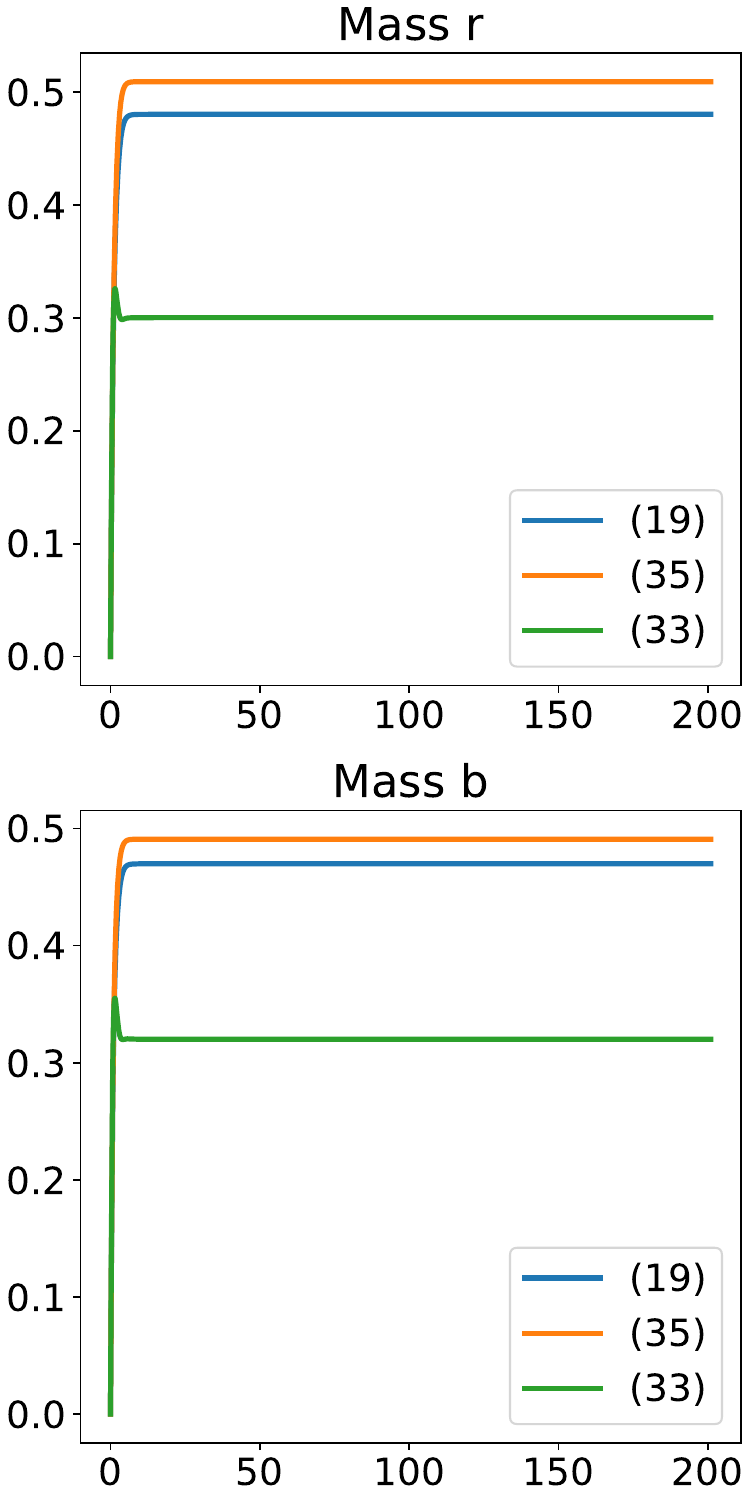}
    }
        \hspace{0.2cm}
    \subfigure[$\alpha_r=0.1$, $\beta_r=0.2$,   $\alpha_b=0.2$,\quad $\beta_b=0.4$]{
    \includegraphics[width=0.275\textwidth]{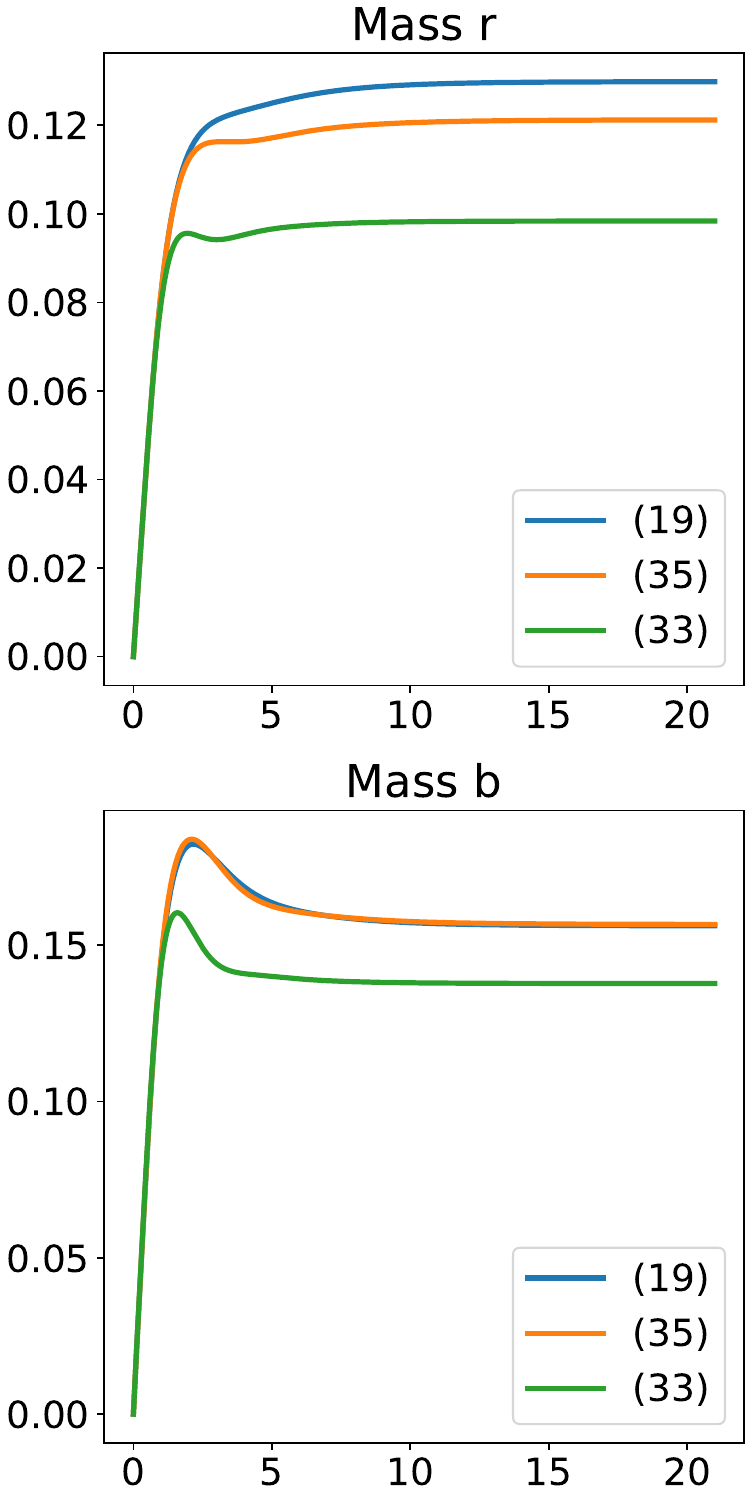}
    }
    \caption{Evolution of the total mass for different flux boundary conditions and with $D_r=D_b = 0.1$ and $\lambda = 0.01$ in all cases.}
        \label{fig:mass}
\end{figure}

In Figure \ref{fig:Regime3} we chose rates above $1/2$ to obtain the maximal-current phase. There, interestingly, it turns our that the dynamics of model \eqref{e:aa_cross_sys} shows a completely different behaviour. This constitutes an interesting starting point for further analytical considerations on the phase behaviour. Figure \ref{fig:mass} displays the evolution of the total mass of the respective species for different in- and outflow rates. We observe that the reaction-diffusion \eqref{model_lattice1D_number_densities} and the lattice based cross diffusion system \eqref{eq:MF_cross_diff} show a similar quantitative behaviour in several in- and outflow regimes, while the cross-diffusion system obtained via asymptotic expansion \eqref{e:aa_cross_sys} behaves only qualitatively similar.
\subsubsection*{Periodic boundary conditions}
\begin{figure}
    \centering
    \subfigure[$t=0$]{
    \includegraphics[width=0.23\textwidth]{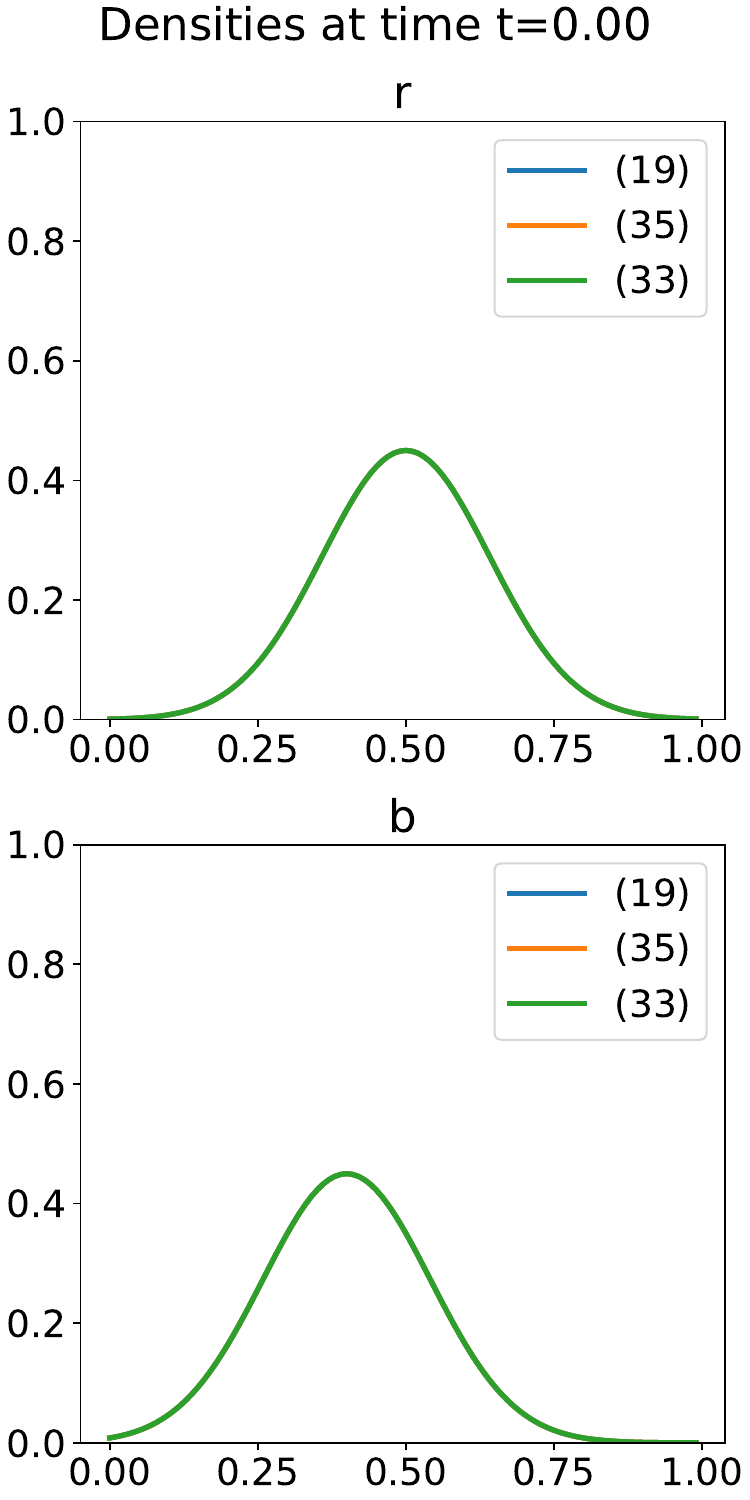}
    }
    \subfigure[$t=0.4$]{
    \includegraphics[width=0.23\textwidth]{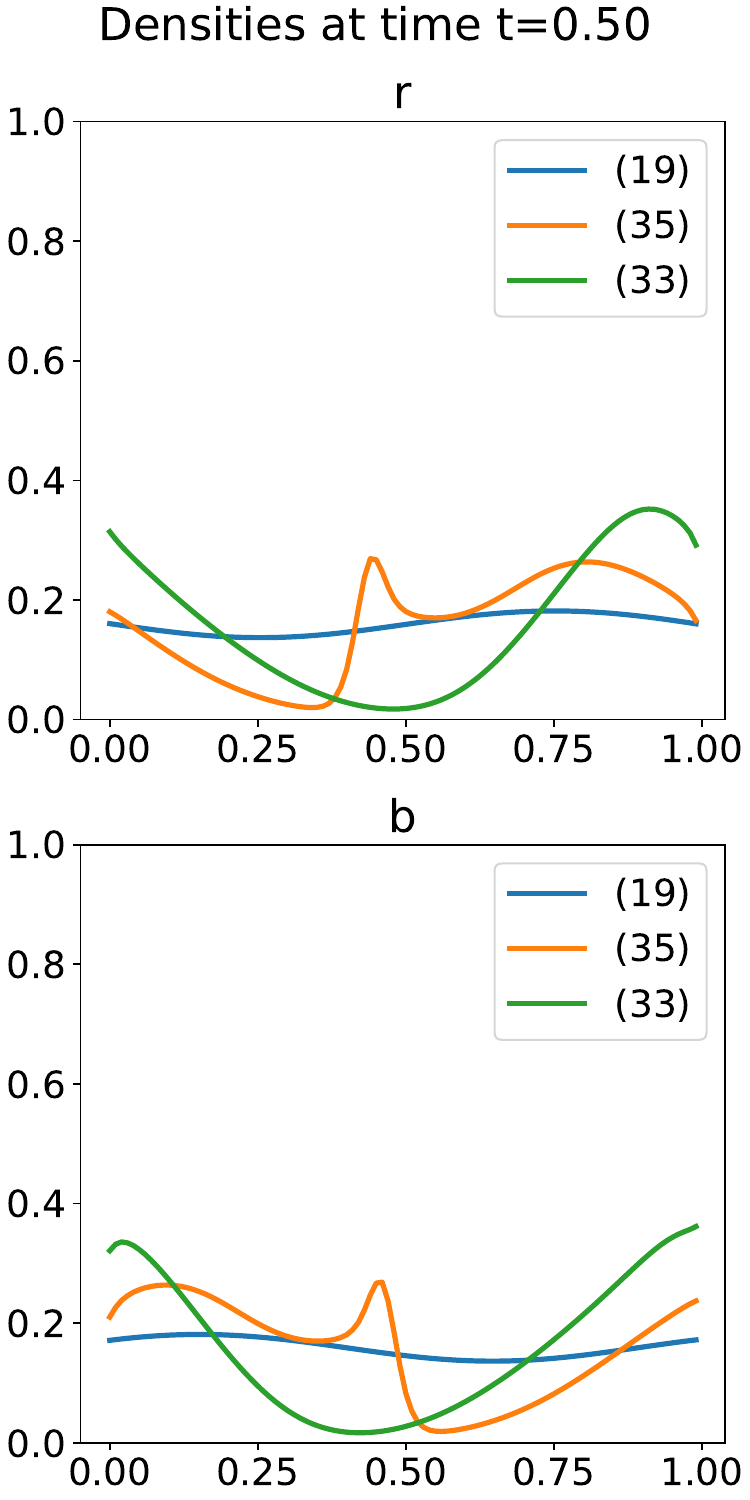}
    }
    \subfigure[$t=1$]{
    \includegraphics[width=0.23\textwidth]{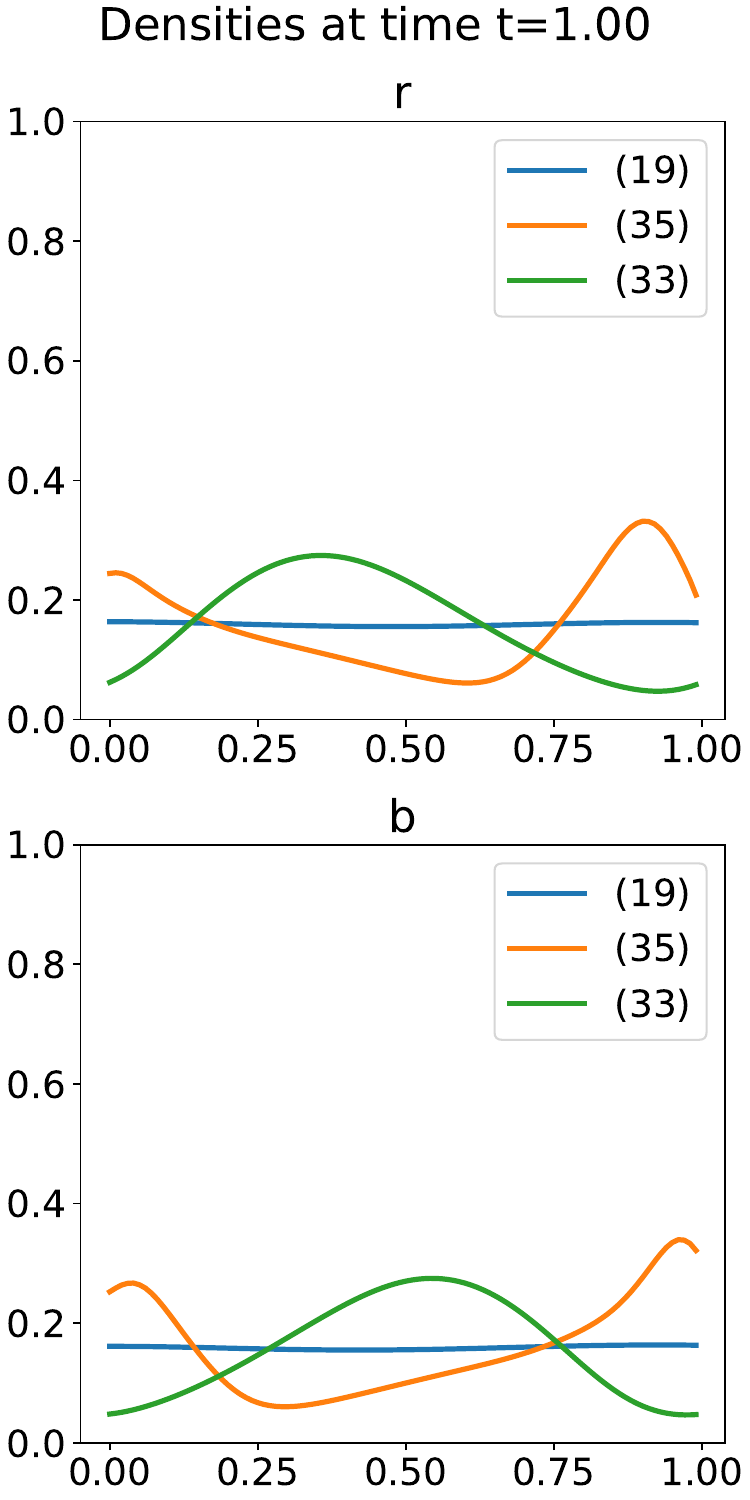}
    }
    \subfigure[$t=3.9$]{
    \includegraphics[width=0.23\textwidth]{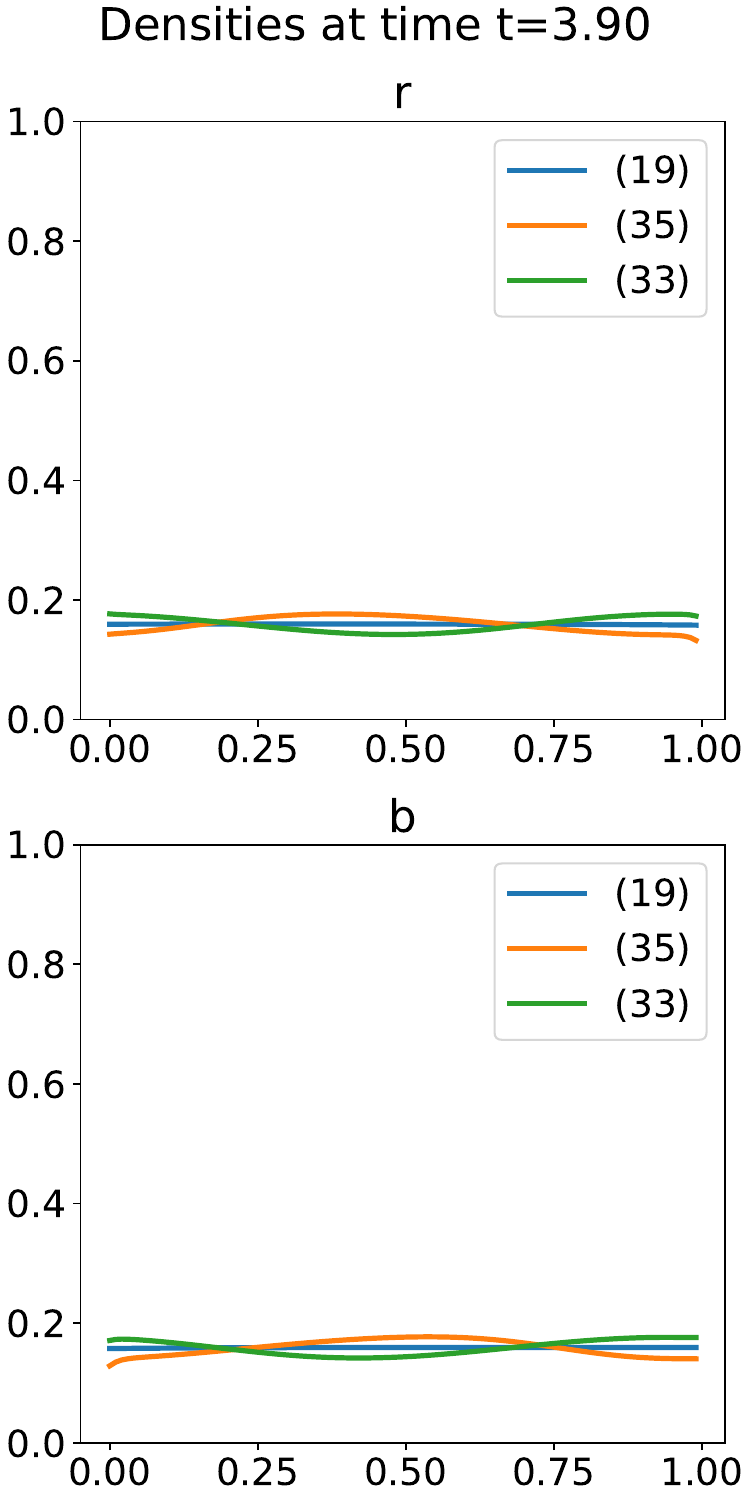}
    }
    \caption{Periodic boundary conditions with $D_r=D_b = 0.01$ and  $\lambda = 0.01$. All models converge to constant stationary solution.}
    \label{fig:periodic}
\end{figure}

For periodic boundary conditions, noting that the velocity is constant, thus periodic, we expect constant stationary solutions whose value is determined by the initial mass. This is indeed observed in Figure \ref{fig:periodic}. However, for earlier times, their dynamics differs substantially, in particular for \eqref{eq:MF_cross_diff}, the influence of cross-diffusion (``jams'') is most pronounced.

\subsubsection*{Confining potential}
\begin{figure}
    \centering
    \subfigure[$t=0$]{
    \includegraphics[width=0.23\textwidth]{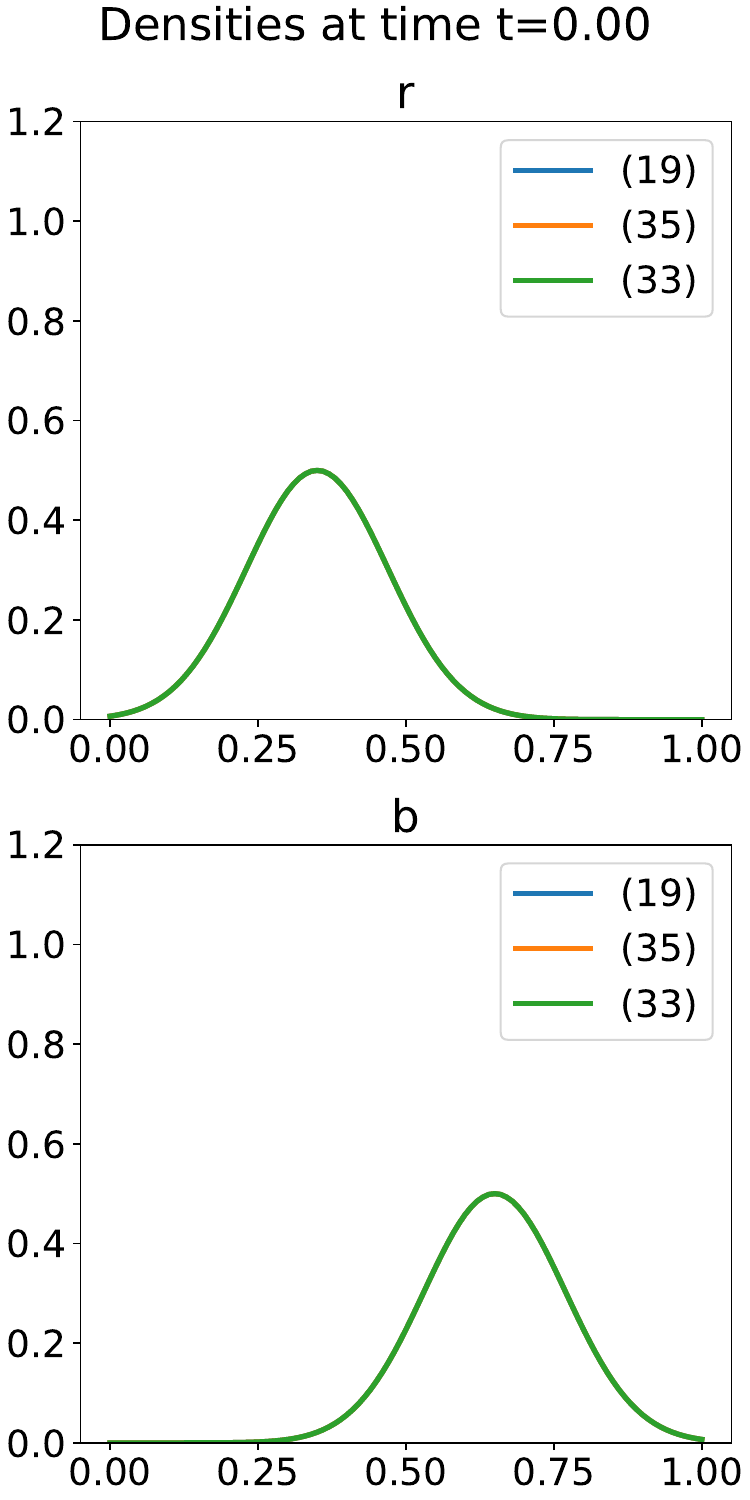}
    }
    \subfigure[$t=0.5$]{
    \includegraphics[width=0.23\textwidth]{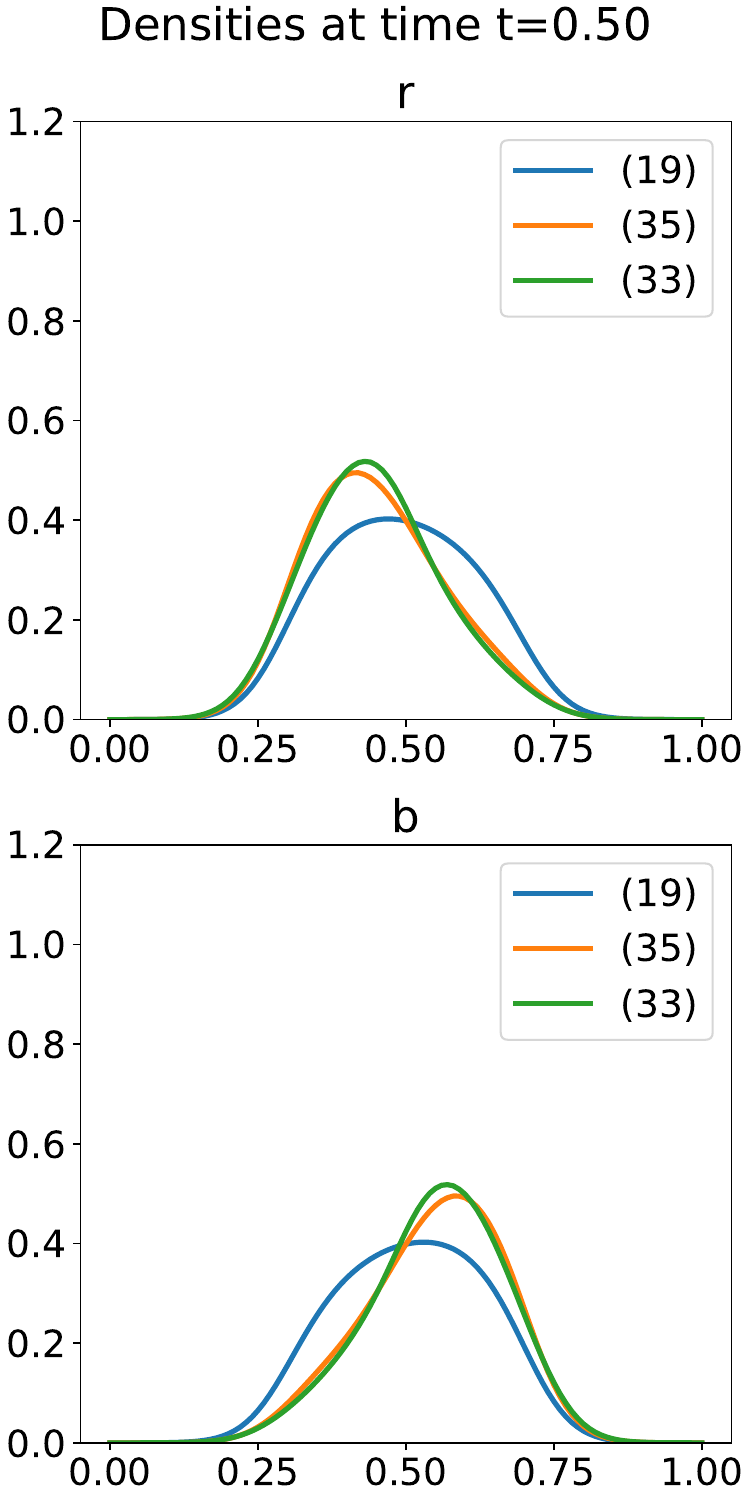}
    }
    \subfigure[$t=2$]{
    \includegraphics[width=0.23\textwidth]{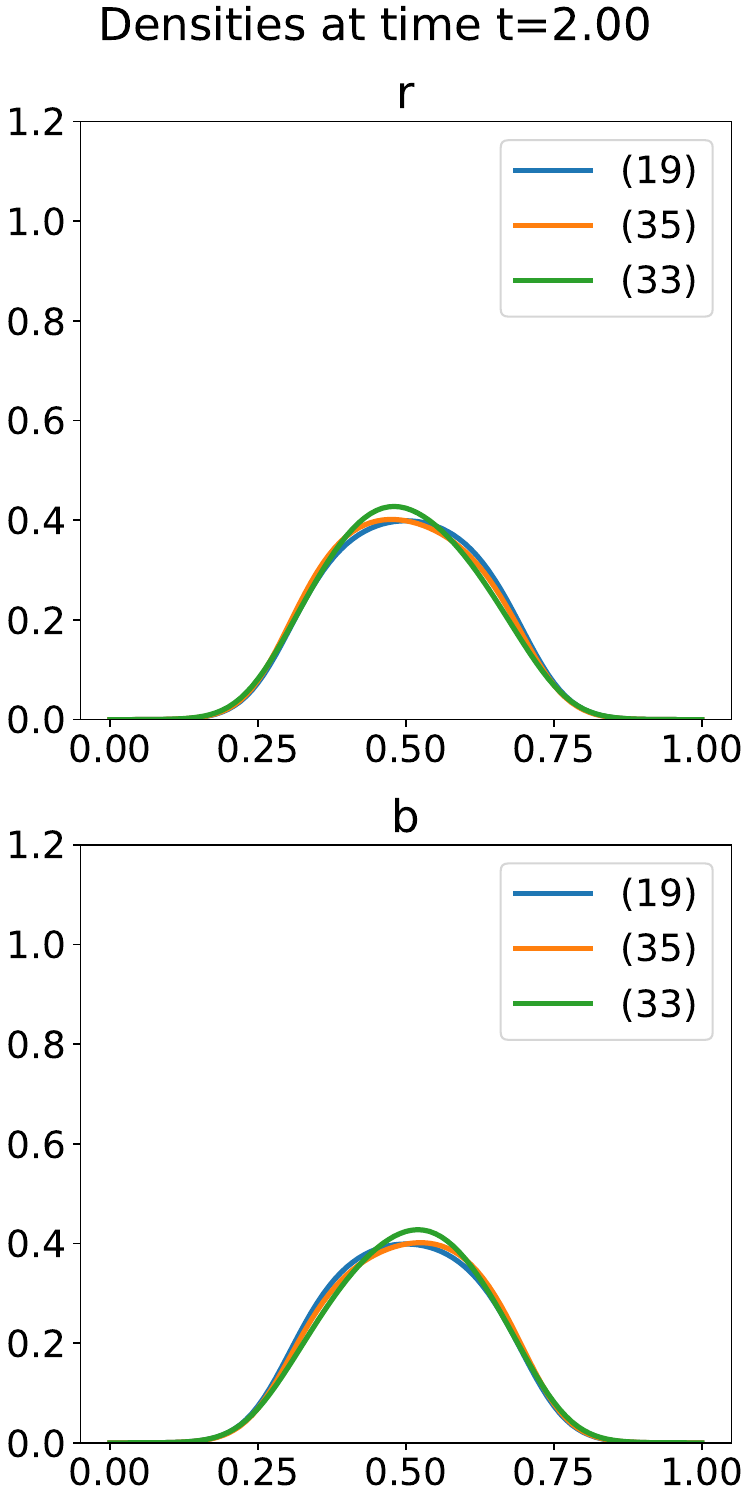}
    }
    \subfigure[$t=10$]{
    \includegraphics[width=0.23\textwidth]{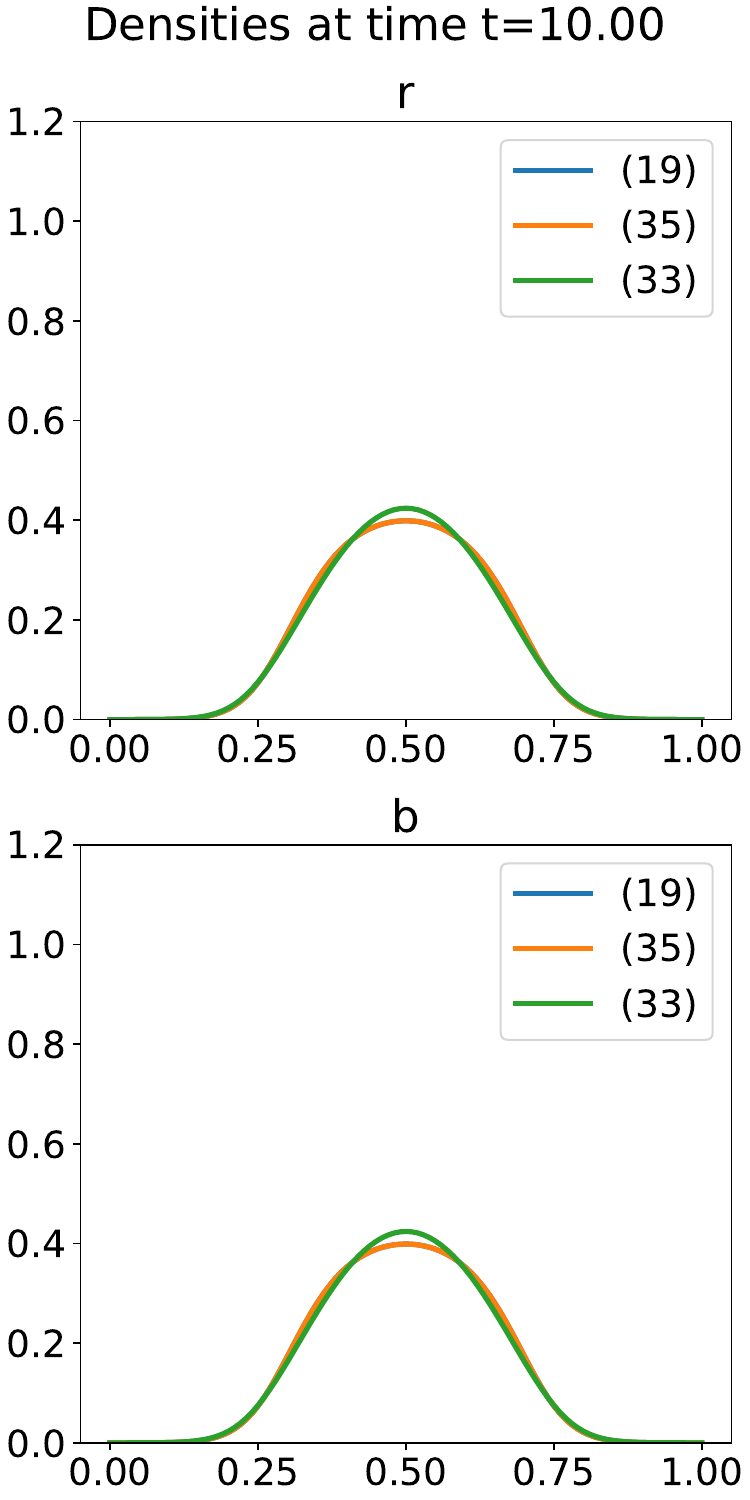}
    }
    \caption{No flux boundary conditions with  $D_r =  D_b = 0.1$, $\lambda = 0.01$ and an confining potential  $V_r = V_b = 5(x-0.5)^2$. }
    \label{fig:confined}
\end{figure}

Finally in Figure \ref{fig:confined}, we consider the situation of no-flux conditions together with a confining potential $V(x) = (x-\frac{1}{2})^2$. Here we observe very similar behaviour for all models, probably due to the fact that the transport term dominates the dynamics. 

\subsection{Two spatial dimensions} \label{sec:numerics2}

In this subsection we reproduce numerical examples in two spatial dimensions from \cite{Bruna:2021tb}. In particular, we show examples of the active continuous model \eqref{model3}, the active hybrid model \eqref{model_hy} and the passive version of the latter, \eqref{eq:MF_cross_diff}, which corresponds to setting $D_R = 0$ in \eqref{model_hy} and choosing an initial condition in angle of the form $\delta(\theta - \theta_1) + \delta(\theta - \theta_2)$ with $\theta_i$ such that $V_r = - (v_0/D_T) \e(\theta_1) \cdot \x$ and $V_b = - (v_0/D_T) \e(\theta_2) \cdot \x$. 
Throughout this subsection we use periodic boundary conditions in the spatial domain $\Omega = [0, 1]^2$, as well as in the angular domain $[0, 2\pi]$ for the active models \eqref{model3} and \eqref{model_hy}. We use the first-order finite-volume scheme of \cite{Bruna:2021tb}, which is based on \cite{Carrillo:2017uq, Schmidtchen:2020wy}. The scheme is implemented in Julia. We use a discretisation with 21 points in each direction and a time-step $\Delta t \le  10^{-5}$ satisfying the CFL condition given by Theorem 3.2 in  \cite{Carrillo:2017uq}. 

\begin{figure}
	\begin{center}
		\includegraphics[width=\textwidth]{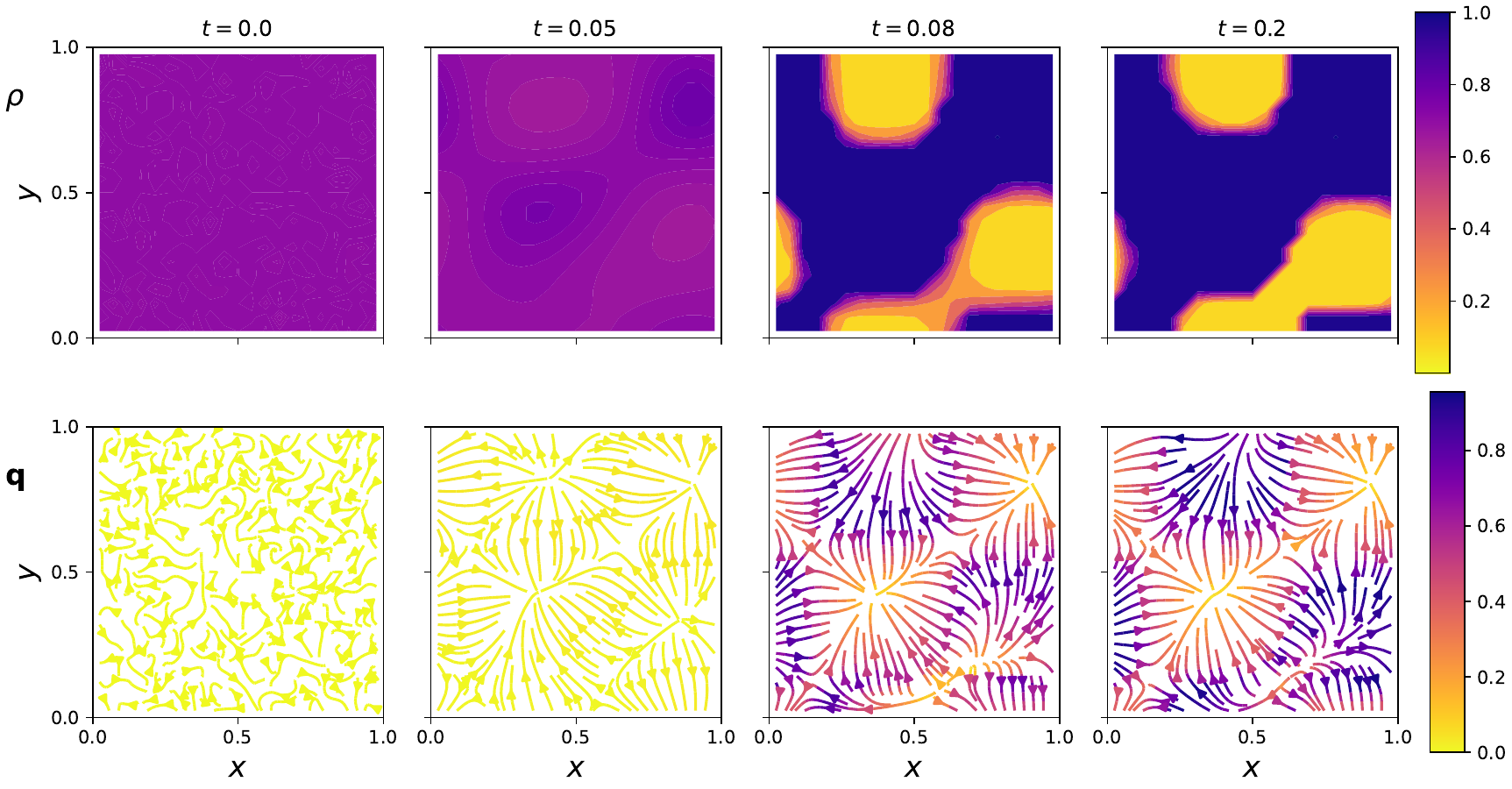}
	\end{center}
	\caption{Hybrid model for active particles \eqref{model_hy} with periodic boundary conditions in $[0, 1]^2 \times [0, 2\pi]$ at different times, starting from a $3d$-random perturbation around the homogeneous solution. The first row shows the total density $\rho(\x,t)$  with mass $\phi$ and the second row  the mean direction $\q(\x,t) $ \eqref{mean_direction}. Parameters used: $D_T = D_R = 1$, $v_0 = 60$, $\phi = 0.7$.}
	\label{fig:model3_70_60}
\end{figure}

\begin{figure}
	\begin{center}
		\includegraphics[width=\textwidth]{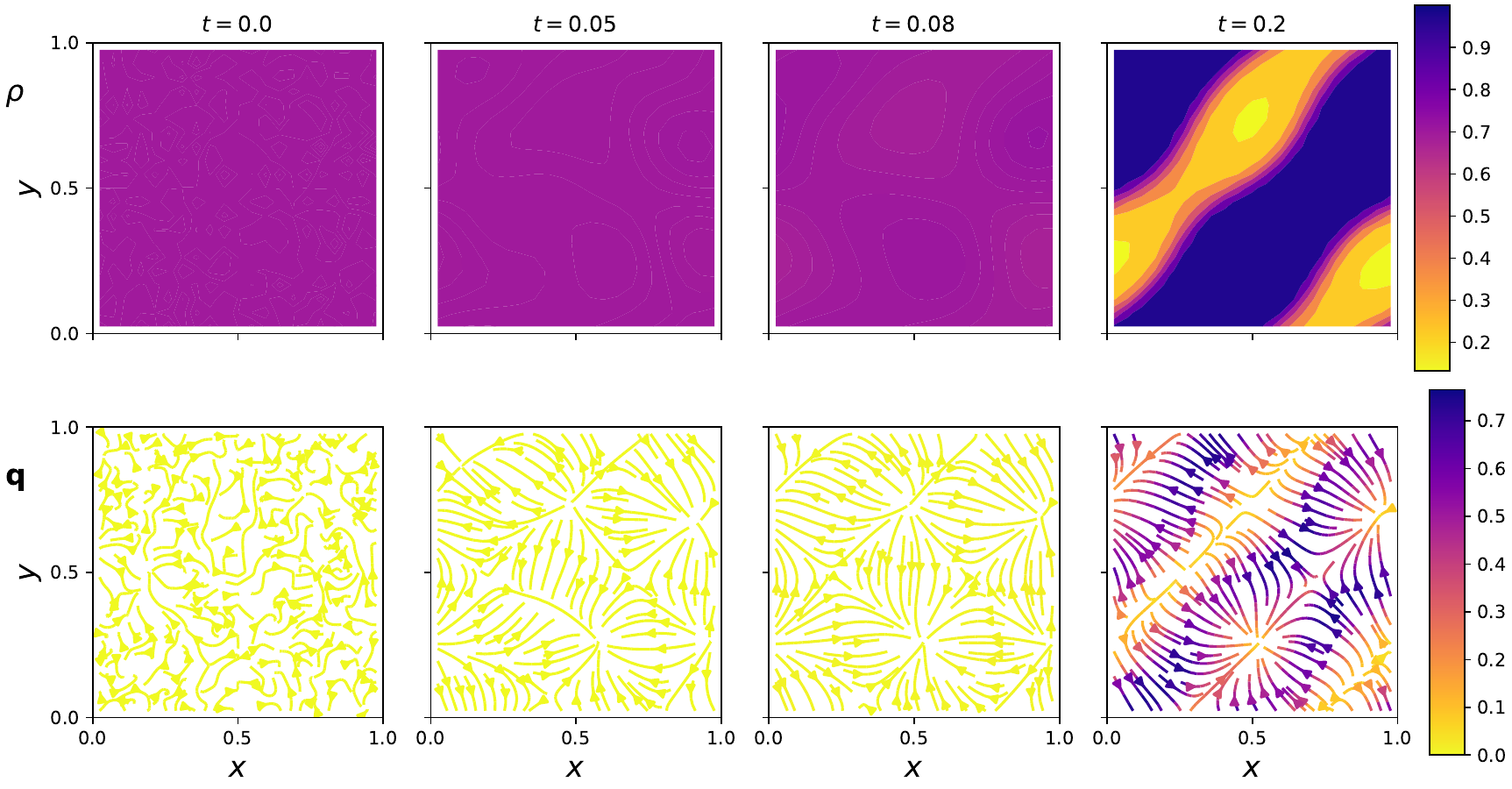}
	\end{center}
	\caption{Hybrid model for active particles \eqref{model_hy} with periodic boundary conditions in $[0, 1]^2 \times [0, 2\pi]$ at different times, starting from a $3d$-random perturbation around the homogeneous solution. The first row shows the total density $\rho(\x,t)$ with mass $\phi$ and the second row  the mean direction $\q(\x,t)$. Parameters used: $D_T = D_R = 1$, $v_0 = 60$, $\phi = 0.7$.}
	\label{fig:model4_70_60}
\end{figure}

Figures \ref{fig:model3_70_60} and \ref{fig:model4_70_60} show the outputs of the two active models \eqref{model3} and \eqref{model_hy} using the same parameters, $D_T = D_R = 1$, $v_0 = 60$, $\phi = 0.7$. In both case, we observe the formation of \emph{motility-induced phased separation} (MIPS), namely a separation into dilute and dense regions and a polarisation of particles in the boundary between these two regions pointing towards the dense region. We show the rescaled spatial density $\rho(\x,t) = \phi \int f \ud \theta$ as well as the mean direction 
\begin{equation}
\label{mean_direction}	
\q(\x,t) := \frac{\int f \e_\theta \ud \theta }{ \int f \ud \theta}.
\end{equation}

\begin{figure}
	\begin{center}
		\includegraphics[width=\textwidth]{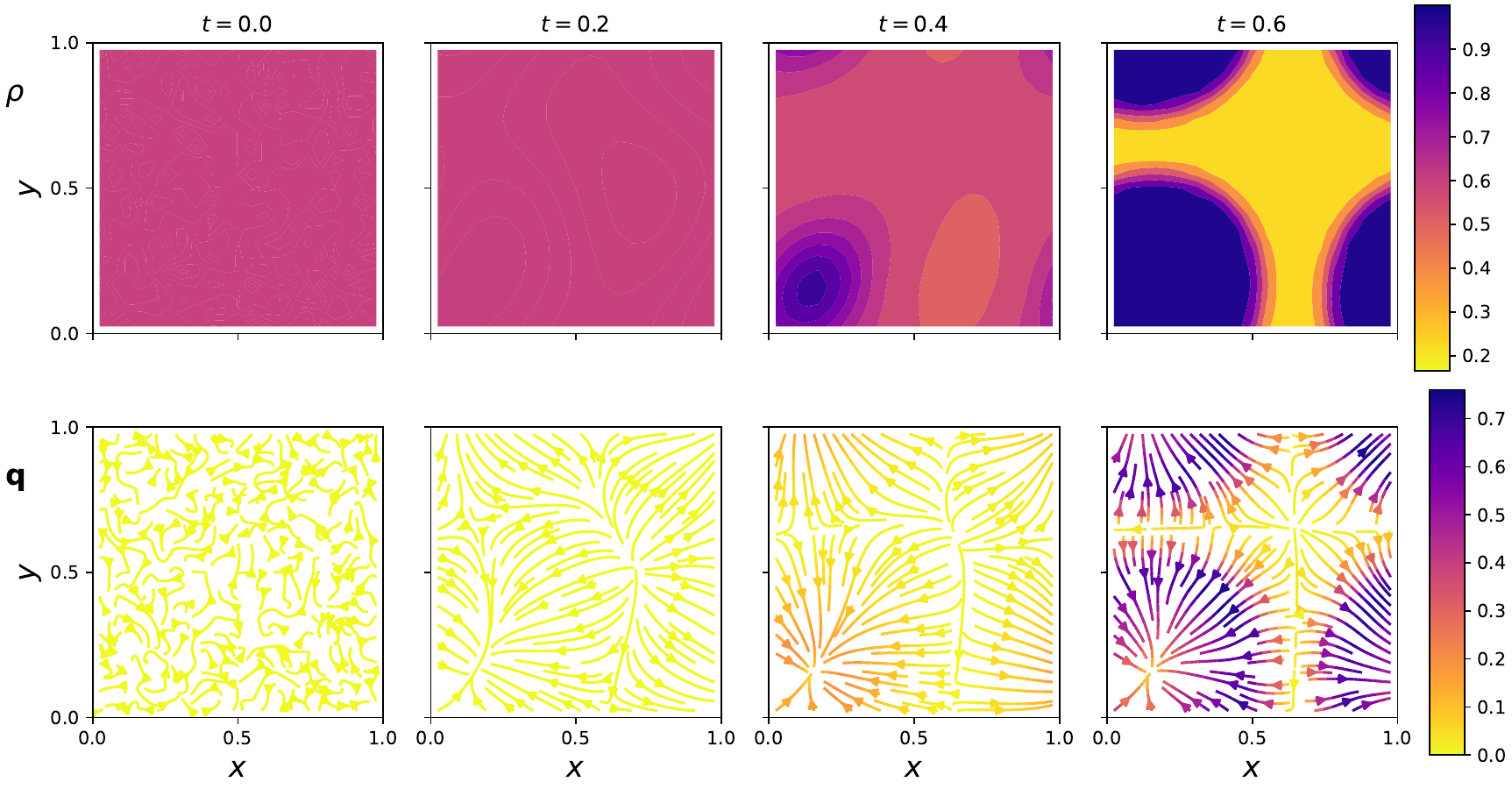}
	\end{center}
	\caption{Hybrid model for active particles \eqref{model_hy} with periodic boundary conditions in $[0, 1]^2 \times [0, 2\pi]$ at different times, starting from a $3d$-random perturbation around the homogeneous solution. The first row shows the total density $\rho(\x,t)$ with mass $\phi$ and the second row  the mean direction $\q(\x,t)$ \eqref{mean_direction}. Parameters used: $D_T = D_R = 1$, $v_0 = 60$, $\phi = 0.6$.}
	\label{fig:model4_60_60}
\end{figure}

\begin{figure}
	\begin{center}
		\includegraphics[width=\textwidth]{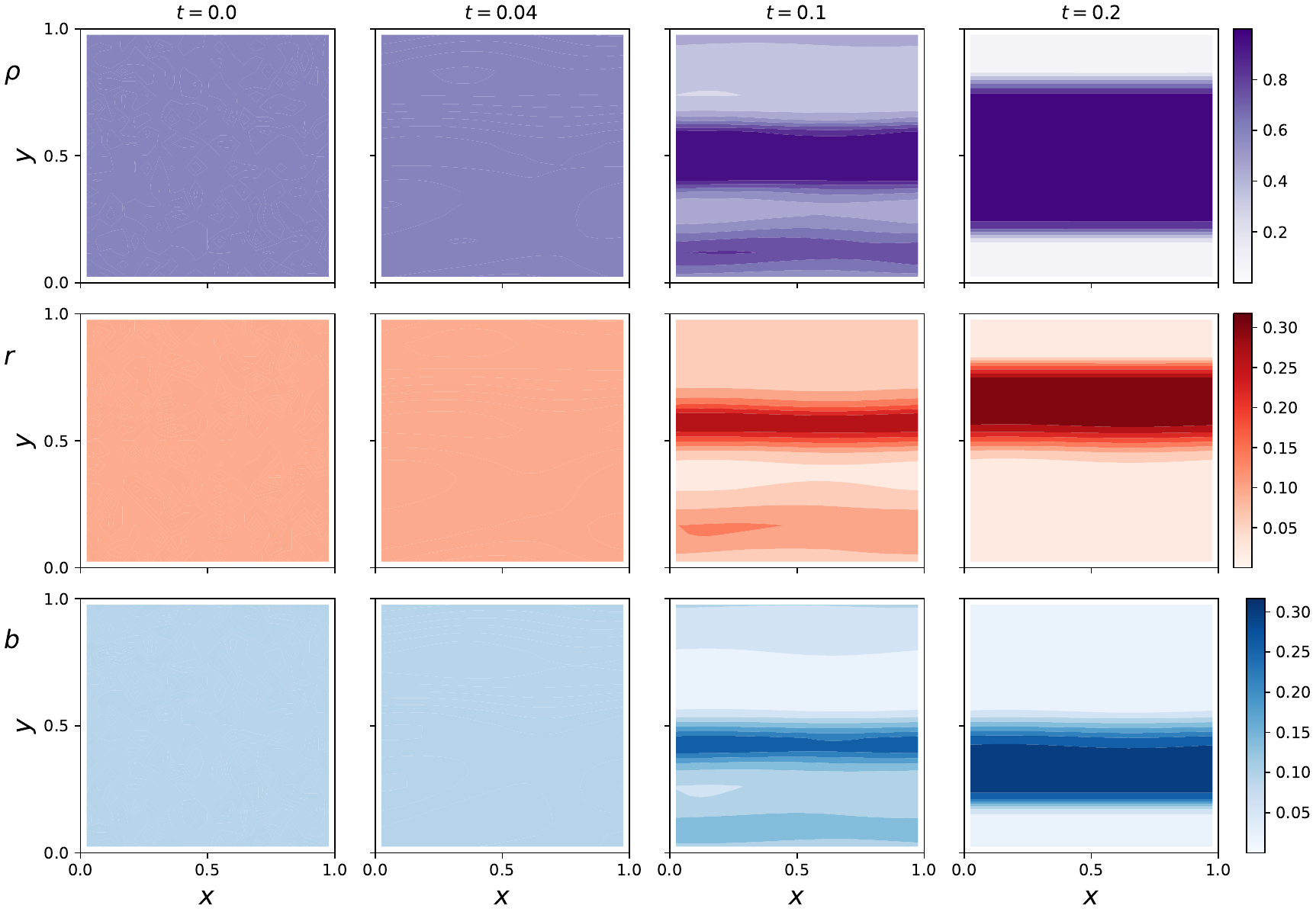}
	\end{center}
	\caption{Discrete model for passive particles \eqref{eq:MF_cross_diff_sys} with periodic boundary conditions in $[0, 1]^2$ at different times, starting from a $2d$-random perturbation around the homogeneous solution. The first row shows the total density $\rho(\x,t) = r + b$ with mass $\phi$, while the densities of the red $r$ and the blue $b$ species are given in the second and third rows, respectively. Parameters used: $D_T = D_R = 1$, $v_0 = 60$, $\phi = 0.6$.}
	\label{fig:model4p_60_60}
\end{figure}

\begin{figure}
	\begin{center}
		\includegraphics[width=\textwidth]{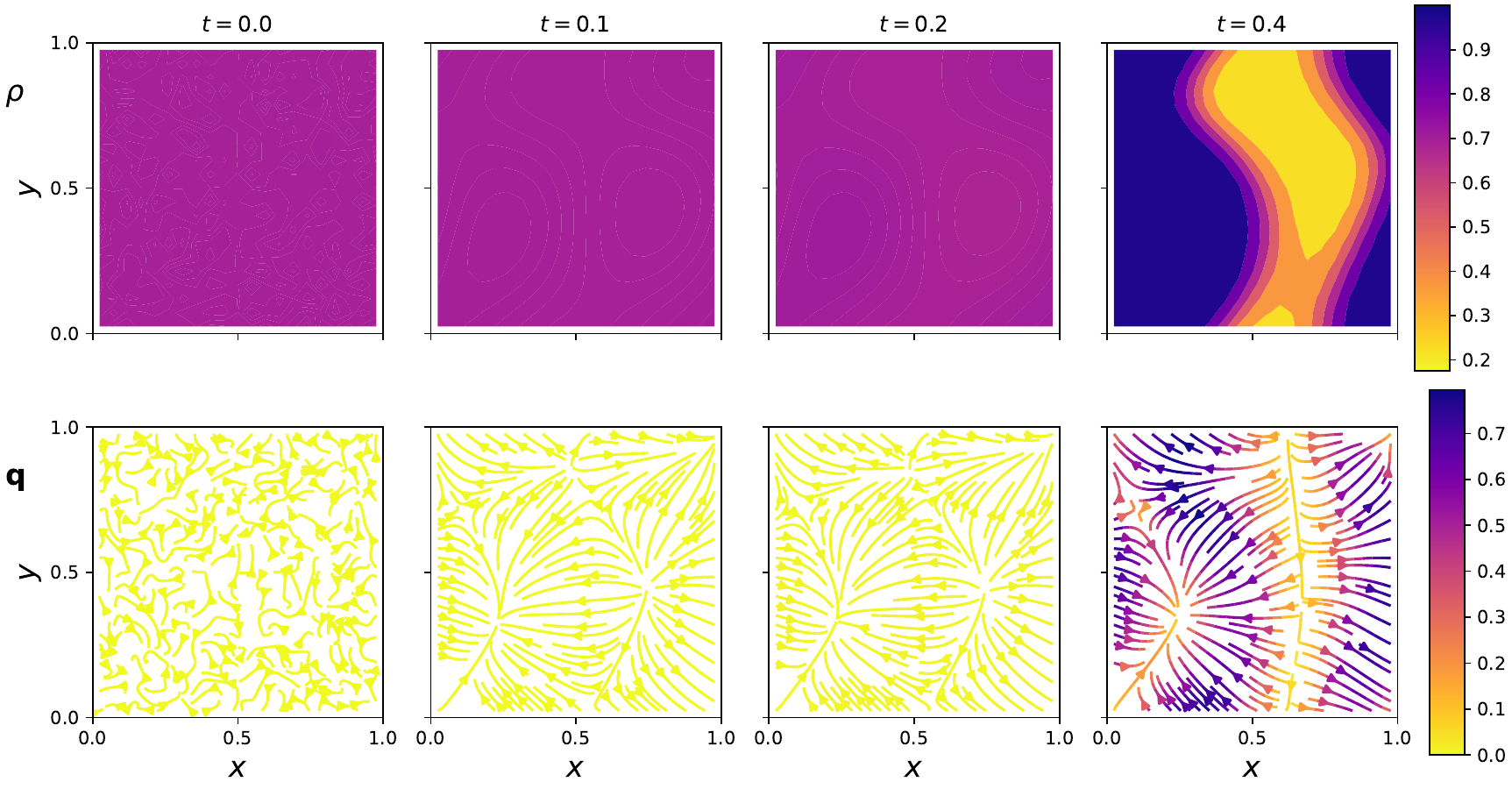}
	\end{center}
	\caption{Hybrid model for active particles \eqref{model_hy} with periodic boundary conditions in $[0, 1]^2 \times [0, 2\pi]$ at different times, starting from a $3d$-random perturbation around the homogeneous solution. The first row shows the total density $\rho(\x,t)$ with mass $\phi$ and the second row  the mean direction $\q(\x,t)$ \eqref{mean_direction}. Parameters used: $D_T = D_R = 1$, $v_0 = 40$, $\phi = 0.7$.}
	\label{fig:model4_70_40}
\end{figure}

\begin{figure}
	\begin{center}
		\includegraphics[width=\textwidth]{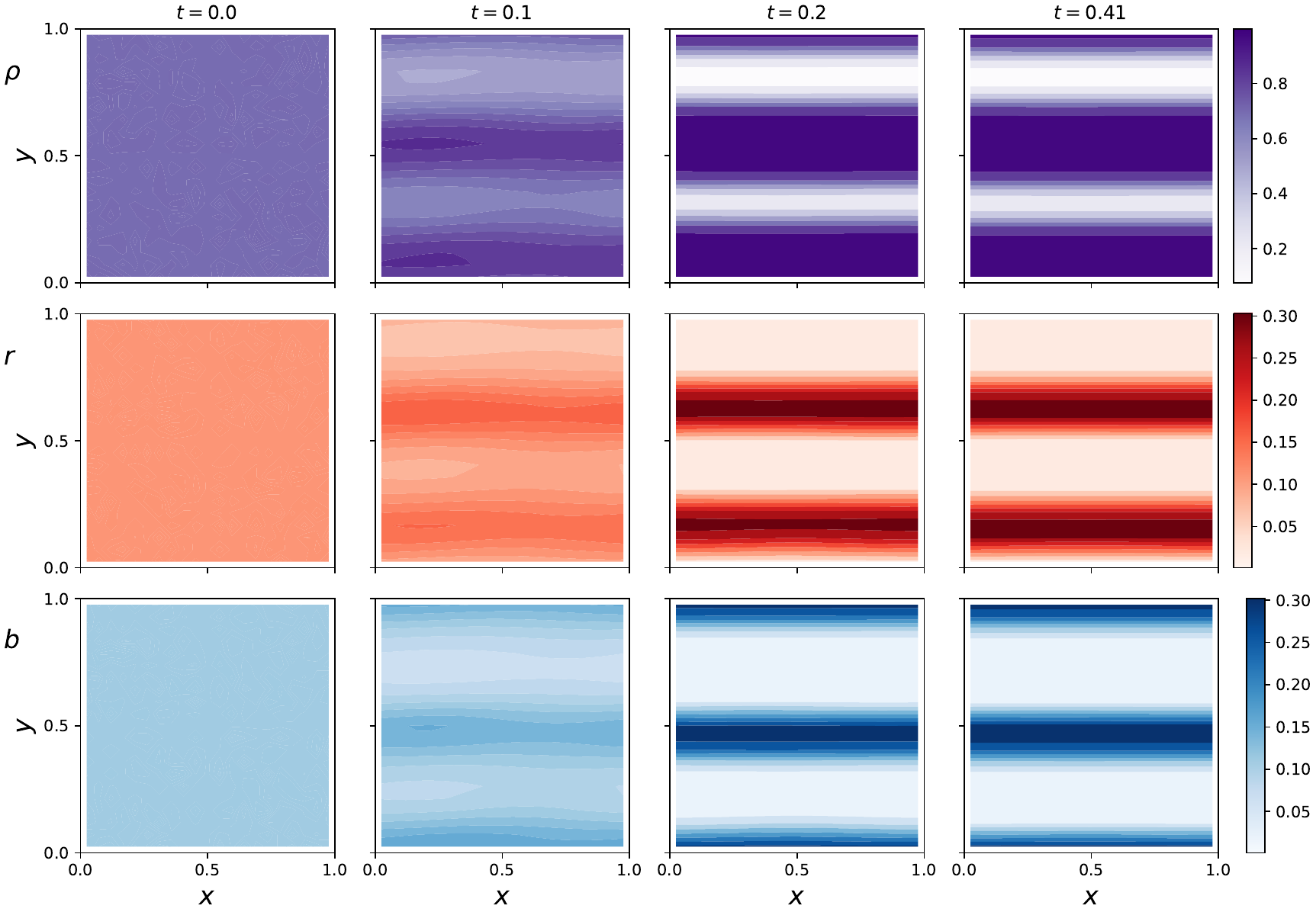}
	\end{center}
	\caption{Discrete model for passive particles \eqref{eq:MF_cross_diff_sys} with periodic boundary conditions in $[0, 1]^2$ at different times, starting from a $2d$-random perturbation around the homogeneous solution. The first row shows the total density $\rho(\x,t) = r + b$ with mass $\phi$, while the densities of the red $r$ and the blue $b$ species are given in the second and third rows, respectively. Parameters used: $D_T = D_R = 1$, $v_0 = 40$, $\phi = 0.7$.}
	\label{fig:model4p_70_40}
\end{figure}

Figures \ref{fig:model4_60_60} and \ref{fig:model4p_60_60} show a comparison between the active hybrid model \eqref{model_hy} and its corresponding passive model \eqref{eq:MF_cross_diff_sys}, which we obtain by setting $D_R = 1$ and a discretisation in angle with only two grid points (which define the two species $r$ and $b$ with respective travel directions $\theta_1 = -\pi/2$ and $\theta_2 = \pi/2$). We observe different types of segregation in each case. In the active case, we observe a blob with high density that is well-mixed in its centre (namely, orientations are uncorrelated as it corresponds to $\q$ small, see bottom right plot in Figure \ref{fig:model4_60_60}). In contrast, in the passive case, in addition to the separation into dilute and dense regions (see first row in Figure \ref{fig:model4p_60_60}), we observe a segregation of the two species within the dense phase: the red particles, which want to move downwards, are met below by a layer of blue particles, which want to move upwards (see second and third rows in Figure \ref{fig:model4p_60_60}). A similar structure in the active model \eqref{model_hy} can be observed if the final pattern is mappable to a one-dimensional pattern, as in the case shown in Figure \ref{fig:model4_70_40} (which corresponds to different values of $\phi$ and $v_0$ and a different initial condition). In this case, ``left''-moving particles concentrate at one boundary and ``right''-moving particles at the other. For this same parameters, the passive model \eqref{eq:MF_cross_diff_sys} displays four instead of two lanes (see Figure \ref{fig:model4p_70_40}). 

\begin{figure}
\begin{center}
	\includegraphics[width=0.45\textwidth]{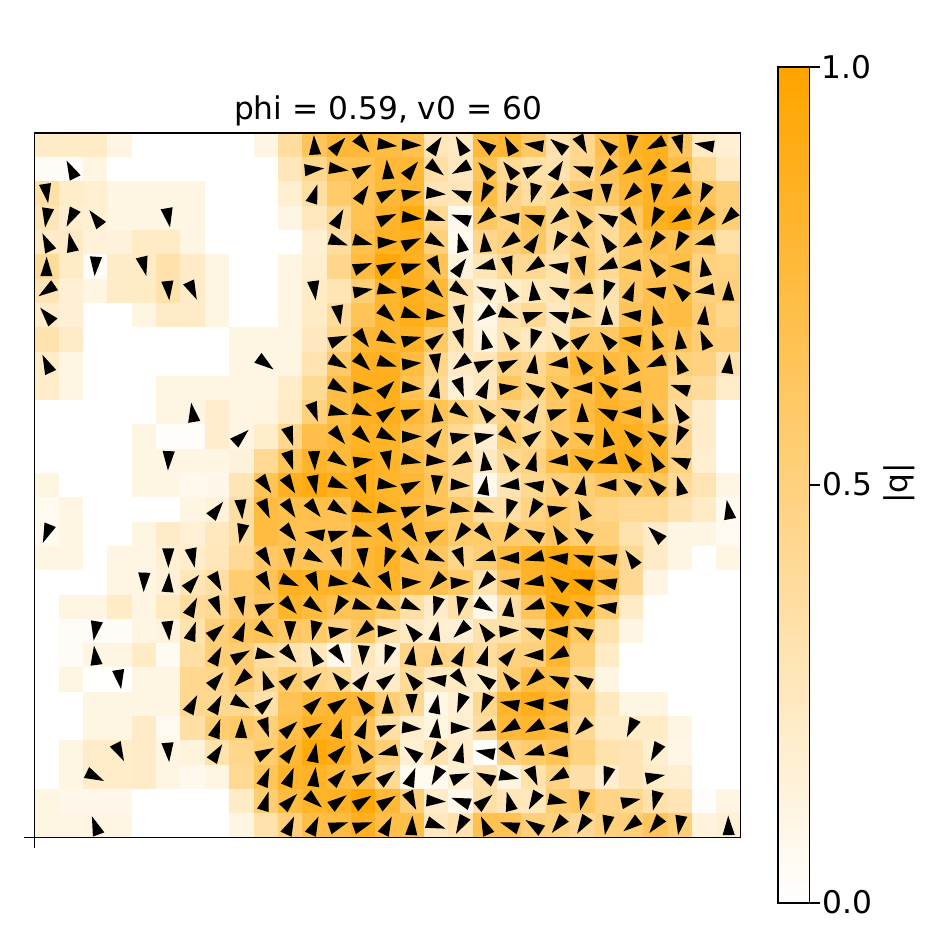}
	\includegraphics[width=0.45\textwidth]{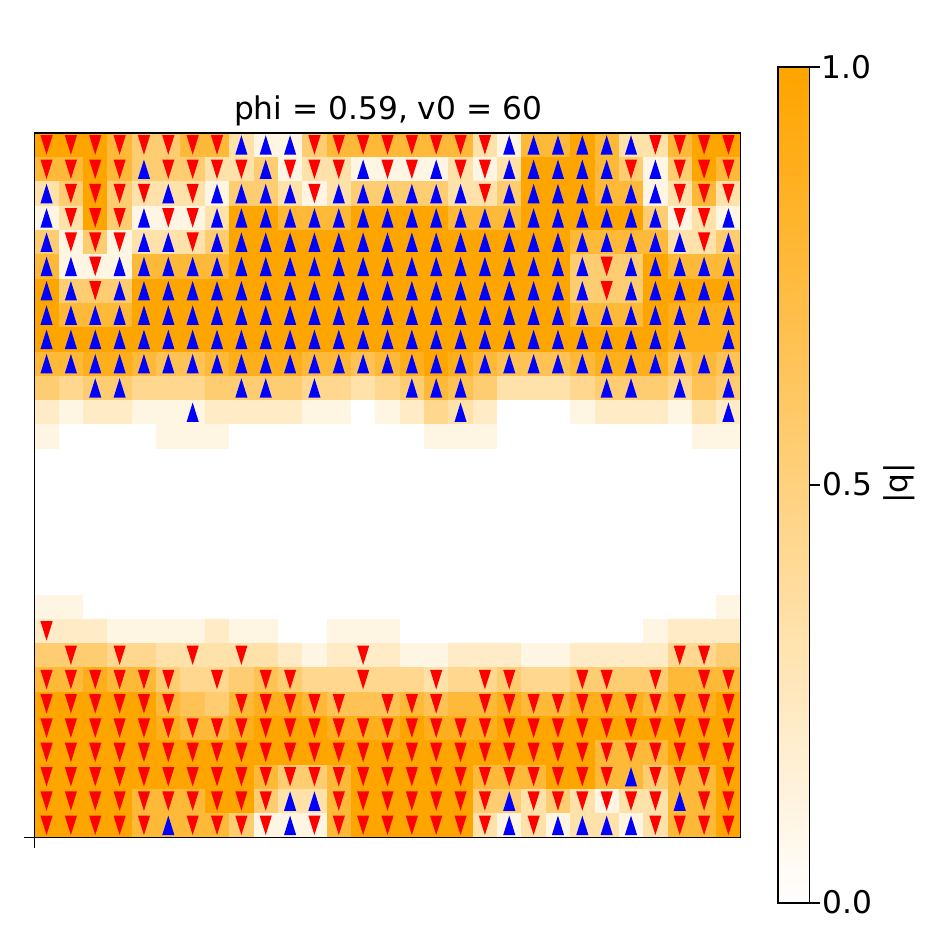}\\
	\includegraphics[width=0.45\textwidth]{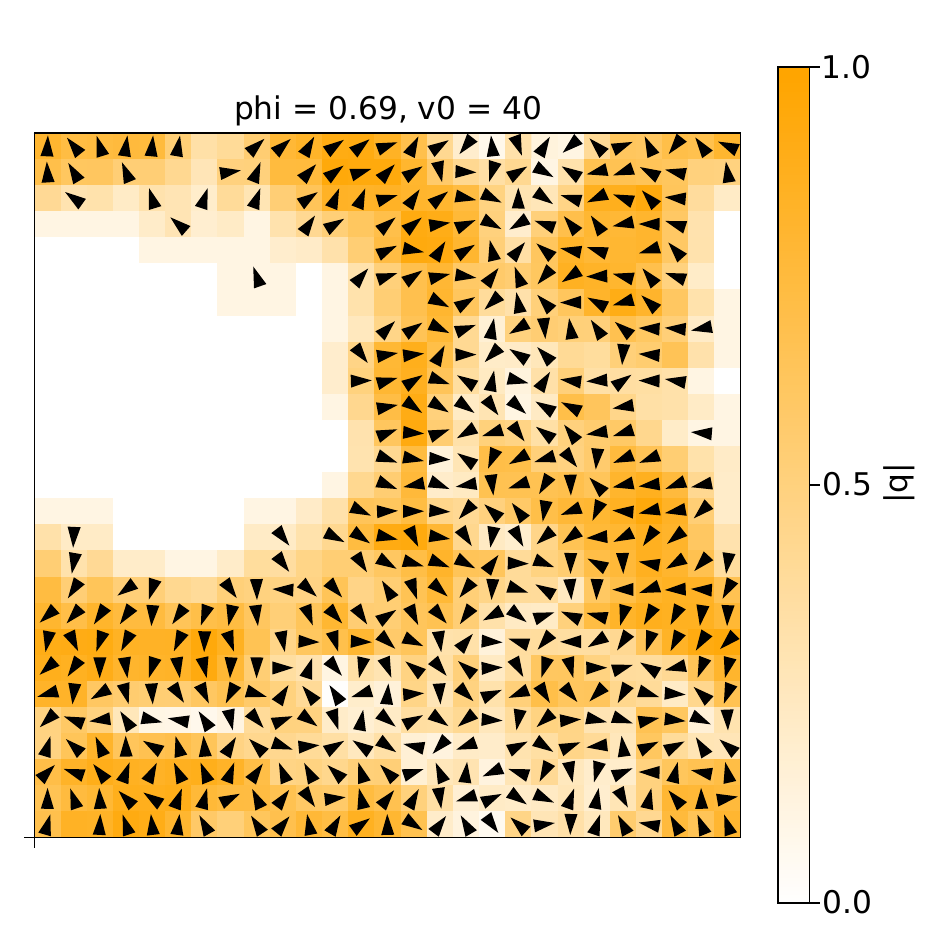}
	\includegraphics[width=0.45\textwidth]{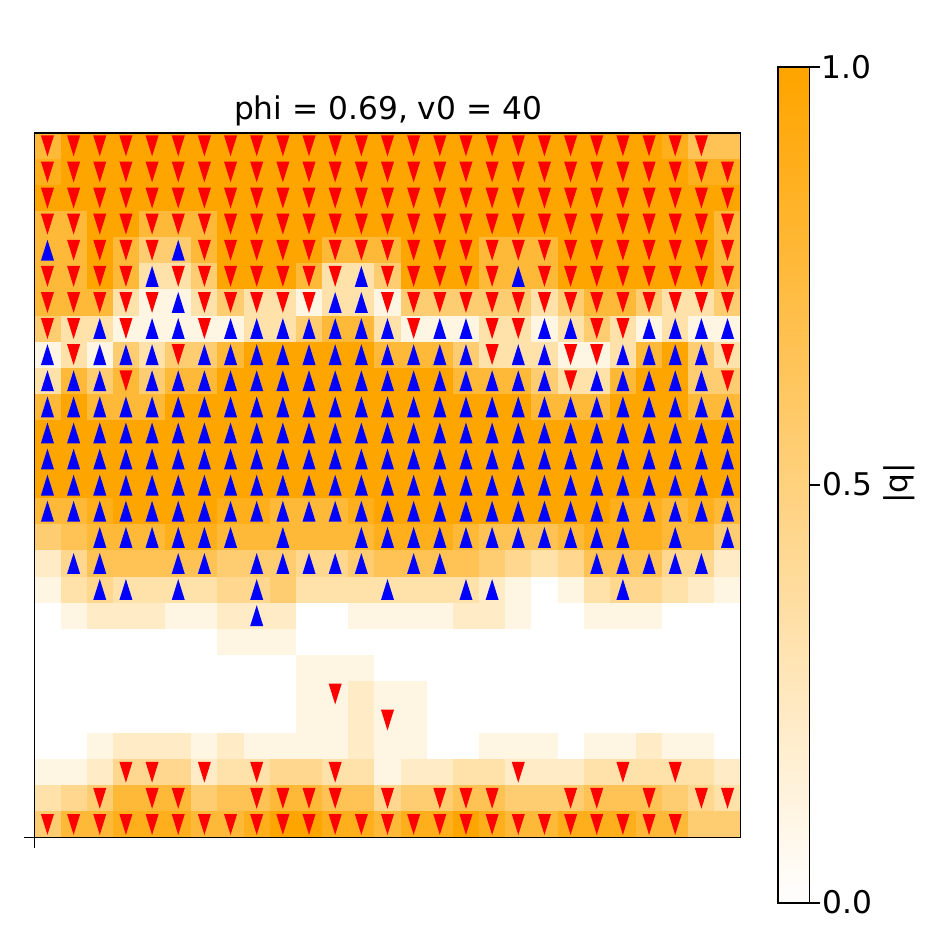}
	
\end{center}
\caption{Example configurations of two microscopic models corresponding to ASEP in position and different rules in orientation. (Left column): hybrid model with Brownian motion in angle with corresponding macroscopic model \eqref{model_hy}. (Right column): passive orientations as given by the initial condition, here either $\theta = \pm \pi/2$, corresponding to the macroscopic model \eqref{eq:MF_cross_diff_sys} for two species of red and blue particles. Snapshots at $T= 0.5$ using fixed time-steps of $\Delta t = 10^{-4}$ and parameters $\phi$ and $v_0$ as indicated above each plot. The colormap shows the strength of the polarisation $|q|$.}
\label{fig:microABM}
\end{figure}

Finally, we show simulation examples of the stochastic models corresponding to the active \eqref{model_hy} and the passive \eqref{eq:MF_cross_diff_sys} macroscopic models. Simulations are performed using the agent-based modelling package Agents.jl \cite{Agents.jl} in Julia and as described in \cite{Bruna:2021tb}.
In both cases, $N$ particles perform an ASEP ((a'), (b) and (c) mechanisms of Subsection \ref{sec:discrete}) on a square lattice  with $M$ lattice sites such that the occupied fraction is $\phi = N/M$. In the former case, particles orientation diffuses with $D_R$ in $[0,2\pi]$, so that the direction of the asymmetric jump process for each particle changes in time. In the latter case, particles are initialised as either red (pointing downwards) or blue (pointing upwards) and their orientations are fixed over time. We observe MIPS in all four cases shown in Figure \ref{fig:microABM}, with the active system displaying either a strip or blob pattern (left column) and the passive system having dilute-dense boundaries and red-blue boundaries running left to right as we had already seen in the PDE simulations (right column). The colormap in the figure shows the absolute value of the mean orientation $\q $ in \eqref{mean_direction} computed using a Moore neighbourhood in each lattice: $|\q| = 0$ in purely isotropic regions (and in empty regions) and $|\q| = 1$ in regions with perfectly aligned particles, which happens within each segregated region of blue and red particles in the passive case and, to a lesser extent, in the boundary between dilute and dense regions in the active case. 

\section*{Acknowledgements}
M.~Bruna was partially supported by a Royal Society University Research Fellowship (grant number URF/R1/180040) and a Humboldt Research Fellowship from the Alexander von Humboldt Foundation.
M.~Burger acknowledges partial financial support by European Union's Horizon 2020 research and innovation programme under the Marie Sklodowska-Curie grant agreement No. 777826 (NoMADS) and the German Science Foundation (DFG) through CRC TR 154 ``Mathematical Modelling, Simulation and Optimization using the Example of Gas Networks'', Subproject C06.
The work of M.-T.~Wolfram was partly supported by the Austrian Academy of Sciences New Frontier's grant NFG-0001. J.-F.~Pietschmann thanks the DAAD for support via the PPP project 57447206.  
\printbibliography

\end{document}